\documentclass[fleqn,usenatbib]{mnras}

\usepackage{newtxtext,newtxmath}
\usepackage[export]{adjustbox}

\usepackage[T1]{fontenc}

\DeclareRobustCommand{\VAN}[3]{#2}
\let\VANthebibliography\thebibliography
\def\thebibliography{\DeclareRobustCommand{\VAN}[3]{##3}\VANthebibliography}
\defcitealias{Koppelman_2021}{K21}
\defcitealias{Kim_2021}{KL21}

\usepackage{graphicx}	
\usepackage{amsmath}	
\usepackage{siunitx}

\title[Reduced proper motion selected halo using Gaia DR3]{Hidden deep in the halo: selection of a reduced proper motion halo catalogue and mining retrograde streams in the velocity space}

\usepackage[symbol]{footmisc}

\author[A. Viswanathan et al.]{Akshara Viswanathan,$^{1}$\thanks{E-mail: astroakshara97@gmail.com (AV)}
Else Starkenburg,$^{1}$
Helmer H. Koppelman,$^{2}$
Amina Helmi,$^{1}$
Eduardo Balbinot,$^{1}$
\newauthor and Anna F. Esselink$^{1}$
\\
$^{1}$Kapteyn Astronomical Institute, University of Groningen, Landleven 12, 9747 AD Groningen, The Netherlands\\
$^{2}$School of Natural Sciences, Institute for Advanced Study, Princeton, NJ 08540, USA\\
}

\date{Accepted 2022 December 25. Received 2022 December 23; in original form 2022 July 22}

\pubyear{2023}

\defcitealias{woudenberg2022characterization}{Woudenberg \& Koop et al., 2022}

\begin{document}
\label{firstpage}
\pagerange{\pageref{firstpage}--\pageref{lastpage}}
\maketitle

\begin{abstract}
The Milky Way halo is one of the few galactic haloes that provides a unique insight into galaxy formation by resolved stellar populations. Here, we present a catalogue of $\sim$47 million halo stars selected independent of parallax and line-of-sight velocities, using a combination of Gaia DR3 proper motion and photometry by means of their reduced proper motion. We select high tangential velocity (halo) main sequence stars and fit distances to them using their simple colour-absolute-magnitude relation. This sample reaches out to $\sim$21 kpc with a median distance of $6.6$ kpc thereby probing much further out than would be possible using reliable Gaia parallaxes. The typical uncertainty in their distances is $0.57_{-0.26}^{+0.56}$ kpc. Using the colour range $0.45<(G_0-G_\text{RP,0})<0.715$ where the main sequence is narrower, gives an even better accuracy down to $0.39_{-0.12}^{+0.18}$ kpc in distance. The median velocity uncertainty for stars within this colour range is 15.5 km/s. The distribution of these sources in the sky, together with their tangential component velocities, are very well-suited to study retrograde substructures.  We explore the selection of two complex retrograde streams: GD-1 and Jhelum. For these streams, we resolve the gaps, wiggles and density breaks reported in the literature more clearly. We also illustrate the effect of the kinematic selection bias towards high proper motion stars and incompleteness at larger distances due to Gaia’s scanning law. These examples showcase how the full RPM catalogue made available here can help us paint a more detailed picture of the build-up of the Milky Way halo.
\end{abstract}




\section{Introduction}

A large body of evidence shows that the assembly of the Milky Way is irrefutably hierarchical. The Galactic halo in particular has a non-linear structure with a vast number of chemical and dynamical stellar streams that allow us to study the formation history of our galaxy. It hosts a range of different substructures – from cold streams \citep[e.g.,][]{shipp2018stellar,malhan2018ghostly,ibata2021charting,martin2022pristine,li2022s} to more diffuse merger events such as the recently discovered major merger Gaia-Enceladus \citep{helmi2018merger,belokurov2018co,koppelman2018one}, the currently disrupting Sagittarius \citep{ibata1995sagittarius}, the Helmi streams \citep{helmi1999debris}, Thamnos \citep{koppelman2019multiple}, Sequoia/I'itoi/Arjuna \citep{myeong2019evidence,naidu2020evidence}, LMS-1/Wukong \citep{yuan2020low,naidu2020evidence}, Cetus \citep{newberg2009discovery,yuan2022complexity} and overdensities such as the Hercules-Aquila cloud, the Virgo Overdensity, and Eridanus-Phoenix \citep{belokurov2006field,balbinot2021}.
The cold stellar streams \citep[see also][for a recent compilation of streams discovered to date]{mateu2022galstreams} are thought to be disrupting or disrupted dwarf galaxies or globular clusters that remain as coherent elongated structures for a long time after the progenitor is fully dissolved before getting phase-mixed \citep[see e.g.,][and references therein]{helmi2008stellar,helmi2020}.
Stellar streams as well as more phase-mixed material that can be discovered due to their coherence in energy and angular momentum \citep[e.g.,][]{ruiz2022substructure} are direct evidence that the Milky Way halo is assembled through several merger events in the past.
These accreted streams allow us to understand the history, formation and evolution of the Milky Way and help probe the Galactic acceleration field and dark matter distribution \citep{koposov2010constraining,ibata2021charting}.
Therefore, the Galactic halo is an interesting playing field in near-field cosmology with a vast number of chemical and dynamical stellar streams that allow us to study the formation history of our Galaxy.

With the release of Gaia early Data Release 3 (EDR3) and Data Release 3 (DR3) \citep{2021,babusiaux2022gaia}, we have full astrometric solutions for 1.468 billion sources out of the total 1.811 billion sources that were mapped over a period of 34 months of data collection. Gaia DR3 also provides radial velocities for more than 33 million sources and BP/RP spectra for more than 220 million sources \citep{katz2022gaia,deangeli2022gaia,andrae2022gaia}. This, in combination with large ground-based spectroscopic surveys such as APOGEE \citep{majewski2017apache,ahumada202016th}, GALAH \citep{buder2021galah+}, SDSS SEGUE \citep{yanny2009segue}, and LAMOST \citep{cui2012large}, has provided the community of Galactic archaeology with exciting new prospects for halo cartography and the study of substructures. Because spectroscopic studies are often limited in their apparent magnitude ranges, kinematic and dynamical studies of the more distant Galactic halo and its substructures often make use of tracers that are intrinsically bright, like red giants \citep[e.g.,][]{naidu2020evidence, chandra2022distant}. Blue Horizontal Branch stars and other bright standard candles such as RR Lyrae are also often employed because they give additional distance information and thereby paint a 3D picture \citep[e.g.,][]{deason2018apocenter, starkenburg2019pristine, wang2022probing}.

Many of the research results produced in studying the halo focus on the Gaia data with three dimensional position and velocity space - called the Gaia 6D sample (see a recent application of Gaia 6D sample in \citet{recio2022gaia}) where we can probe the Integrals of Motion space to study the assembly of the halo \citep{helmi1999building,helmi2000mapping}. However, this 6D sample makes up only a small part of the entire Gaia sample without line-of-sight velocities - called the Gaia 5D sample. Only a sub-sample of 2.4\% of the sources out of the 78\% of the sources with astrometric solutions have Gaia line-of-sight velocities. Additionally, if one is seeking to add distance information, one is hampered by the fact that most of the stars in Gaia DR3 suffer from poor parallaxes. Only about $9.5\%$ of the stars in Gaia DR3 have \texttt{parallax$\_$over$\_$error} $>5$ (allowing 20\% error in distances) i.e., only one-tenth of the humongous Gaia DR3 can be used to study sources with precise distance measurements derived from Gaia parallaxes.

Much of the wealth of Gaia DR3's 1.8 billion stars thus lies in the 5D sample with accurate photometry and astrometry, most of which is comprised of low luminosity low mass main sequence stars. Although it comes with clear limitations - such as the lack of line-of-sight properties, poor parallax as well as unconstrained parallax offset for distances beyond 2 kpc, and the fact that the volume (in terms of distances) of stars that can be probed is smaller than that of giants - there is much benefit in mining these stars effectively for Galactic Archaeology purposes. The \texttt{STREAMFINDER} algorithm is a clear example of successful mining of (part of) this large dataset, as it has been finding and characterising many coherent stellar streams in the halo \citep{ibata2021charting}. 

Main sequence stars are moreover very useful tracers to use in addition to brighter stars as they make up the majority of the stellar halo. The low-mass end of the main sequence is very long-lived and these stars preserve the imprint of their birth materials in the atmosphere better than more evolved stars. Most importantly, they vastly outnumber the brighter stars, making them useful to probe very small substructures, or faint surface brightness features. 

In this work, we aim to select halo main sequences stars using the Gaia 5D sample. We explore the selection of Galactic halo main sequence stars out to $\sim21$ kpc using Gaia DR3 proper motions and photometry, which derives the reduced proper motion (see \citealt{1972ApJ...173..671J,gould2007investigation,10.1111/j.1365-2966.2009.15391.x} for more information about the reduced proper motion and its usage in various contexts). A similar selection method was recently used by \citet{Kim_2021} to select half a million local halo main sequence stars, but they restricted their sample to stars out to 2 kpc using the parallax information from Gaia. Here, we instead follow the same method as explored on Gaia DR2 data by \citet{Koppelman_2021} and we include stars without good parallax information. As with the release of Gaia DR3, the limiting Gaia G-magnitude increased to $\sim$22 with the survey being complete up to G<19 \citep{fabricius2021gaia}, we see that our selection method increases the catalogue size five times compared to \citet{Koppelman_2021}. The simple colour-magnitude relation for the main sequence stars works in our favour to fit very reliable photometric distances to these stars, allowing many avenues for exploration.

In this paper, we explore the study of fainter counterparts of retrograde stellar streams by using the binned velocity space of the catalogue. We furthermore improve distance information to candidate stellar stream stars by folding in available metallicity information.

This paper is laid out as follows: In Section \ref{2}, we introduce the reduced proper motion, explain the selection method, photometric distance fitting, catalogue description and distance validation. Binned velocity space is introduced in Section \ref{3}. Selection, validation and the improvement of photometric distances using metallicity information of retrograde stellar streams GD-1 and Jhelum are covered in subsections \ref{3.2} and \ref{3.3} respectively. In subsection \ref{3.4}, we show how a combination of Gaia's scanning law and our high proper motion selection provides us with a complementary view of the Sagittarius stream.
Section \ref{4} presents the summary of our results, future works, synergies with different datasets and potential science cases for the catalogue. 

\begin{figure}
	\includegraphics[width=\columnwidth]{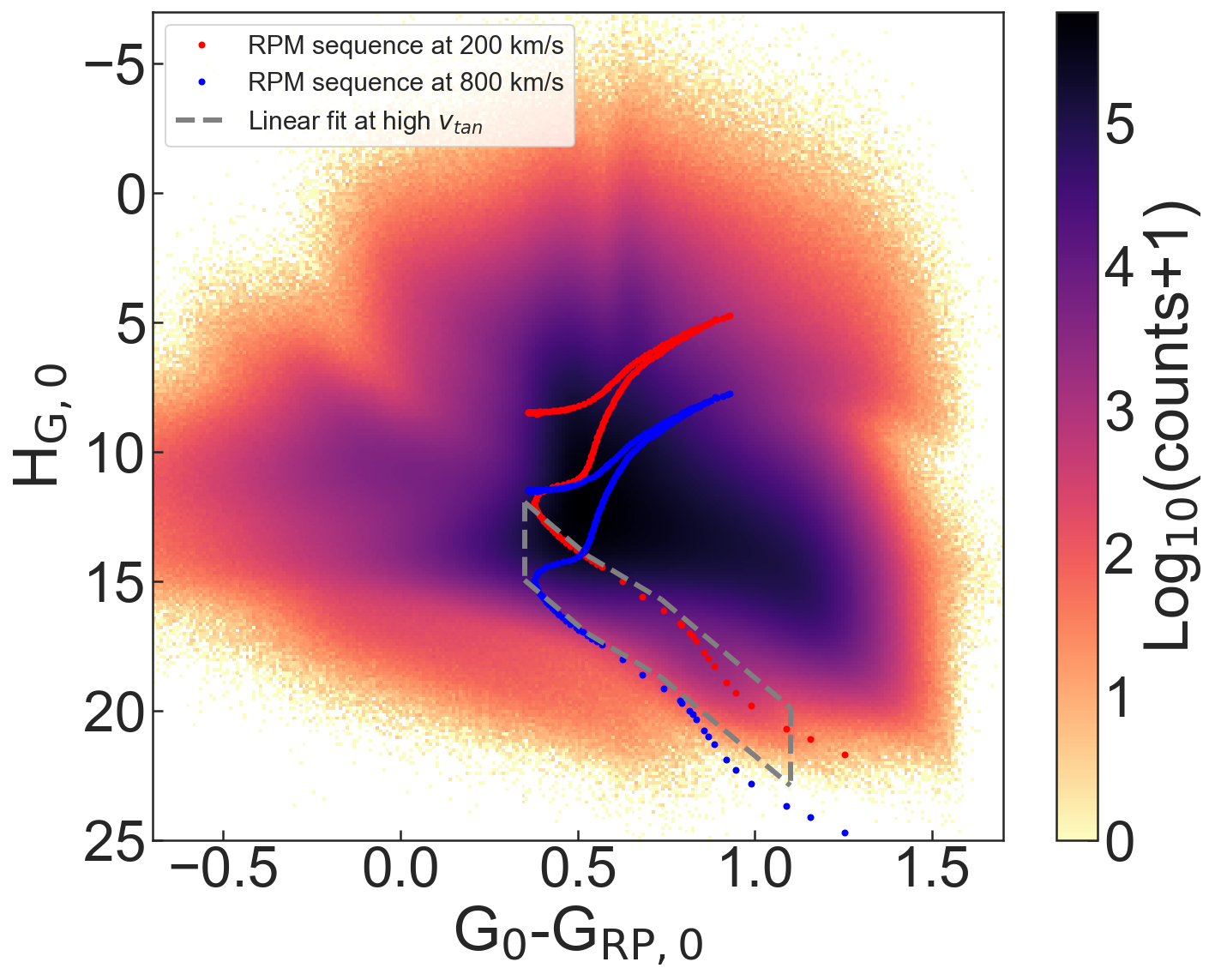}
    \caption{The RPM Diagram for all the sources in Gaia DR3 5D sample that pass astrometry and photometry quality cuts (see text). The high density region consists of mostly disc stars (of $\sim$99\% of the stars). The area inside the polygon enclosed by grey dashed lines represents the tentatively selected halo sources at these high tangential velocities vertically bounded by the main sequence colour range. A -1.6 [M/H], 12 Gyr age isochrone (where absolute magnitude is converted to reduced proper motion parameter) at these higher tangential velocities is over-plotted in red and blue to show that the selections we make correctly picks up halo main sequence stars}
    \label{fig:rpm}
\end{figure}

\begin{figure*}
    \centering
	\includegraphics[width=\textwidth]{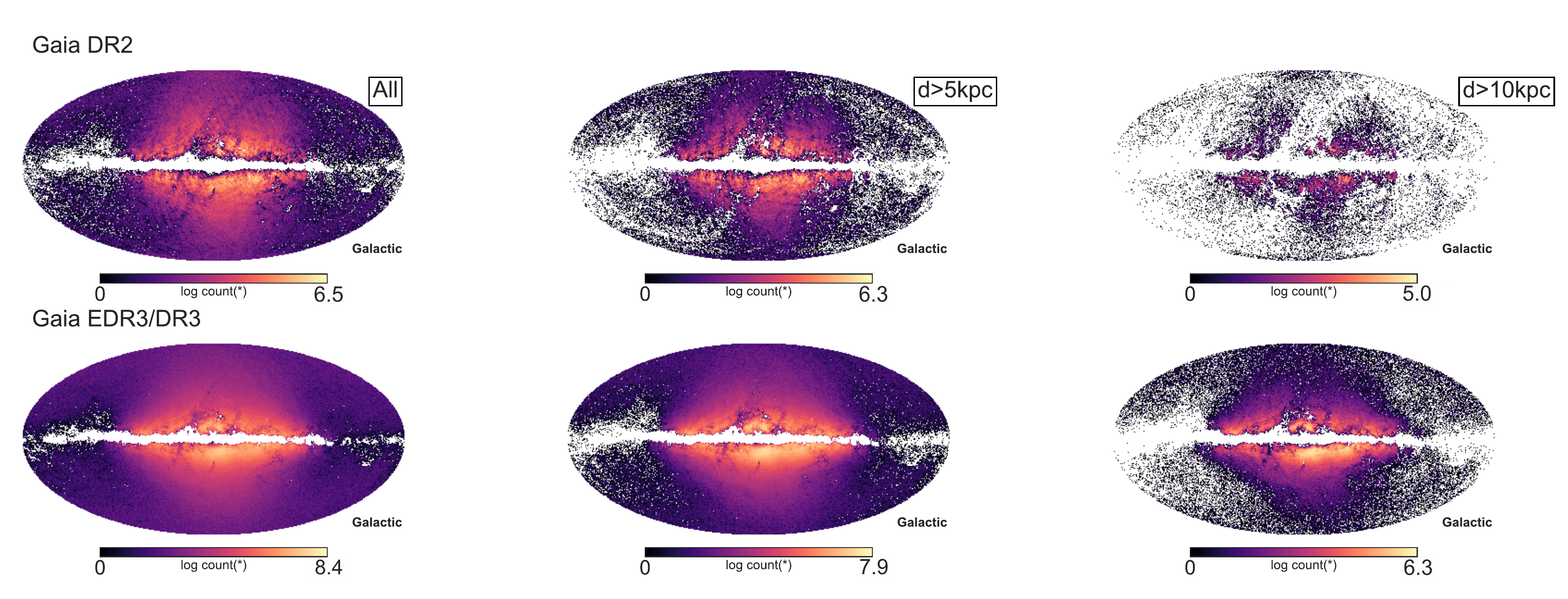}
    \caption{Mollweide map of RPM selected halo stars in Galactic coordinates. Top panels: RPM selected halo using Gaia DR2; All sources (left), sources with heliocentric distances greater than 5 kpc (middle), sources with heliocentric distances greater than 10 kpc (right). Bottom panels: RPM selected halo using Gaia DR3; All sources (left), sources with heliocentric distances greater than 5 kpc (middle), sources with heliocentric distances greater than 10 kpc (right)}
    \label{fig:full-sky-sample}
\end{figure*}

\section{Reduced Proper Motion selected halo sample}\label{2}
In this paper, we use the Gaia data release 3 \citep{babusiaux2022gaia} astrometry and photometry to select halo main sequence stars.
\subsection{High tangential velocity stars in halo orbits}\label{2.1}
The process of selecting main sequence stars on halo-like orbits in Gaia DR2 \citep{2018} using reduced proper motion was explained in \citet{Koppelman_2021} (hereafter \citetalias{Koppelman_2021}) using just the Gaia proper motion (astrometry) and photometry information, the combination of which renders reduced proper motion. 

The (Gaia G-band) reduced proper motion is given by the following equation(s): 
\begin{equation}
    H_\text{G,0}=m_\text{G,0}+5\log\mu-10
	\label{eq:rpm1}
\end{equation}
which is equivalent to:
\begin{equation}
    H_\text{G,0}=M_\text{G,0}+5\log\frac{v_\text{tan}}{4.74057}
	\label{eq:rpm2}
\end{equation}

where $\mu$ is the total reflex corrected proper motion in mas/yr defined as $\sqrt{(\mu_\text{RA}^*\cos(\text{Dec}))^2+\mu_\text{Dec}^*{2}}$ provided by Gaia, $m_\text{G,0}$ is the apparent Gaia G-magnitude after applying extinction correction, $M_\text{G,0}$ is the absolute G-magnitude and $v_\text{tan}$ is the tangential velocity in km/s of the source that is proportional to the product of distance in kpc and proper motion in mas/yr by a factor of $4.74057$ which is the result of unit conversions to obtain the velocity in terms of km/s. 

Using the relation between equation (\ref{eq:rpm1}) and (\ref{eq:rpm2}), it can be seen that we are able to select high tangential velocity stars in the reduced proper motion versus colour diagram (hereafter, the RPM diagram). This is because, main sequence stars follow a simple colour-absolute magnitude relation and therefore, at a fixed $v_\text{tan}$, a simple RPM colour relation. 

Looking at the relation in equation (\ref{eq:rpm2}), we can see that the RPM diagram simply mimics the colour-absolute magnitude diagram if we select stars with similar tangential velocities. Groups of stars that belong to the same substructure with small dispersion in tangential velocity will look similar in the RPM and HR diagram. The $v_\text{tan}$ for disc stars are small, which means that clean samples of halo stars can be easily selected by selecting high $v_\text{tan}$ stars. It should be noted that such kinematically selected halo samples disfavour sources with low tangential velocities (in turn, small proper motions) and high line-of-sight velocities. While such a selection of halo stars thus might be very clean, it is not complete, and it has a dynamical selection bias. 

The dependence of $H_G$ on both absolute magnitude and tangential velocity brings another useful property: we can safely assume that the stars living in regions of the RPM diagram where main sequence stars with high tangential velocities are selected are indeed main sequence stars. If they were on the (much more luminous) giant branch instead, their tangential velocities derived from these proper motions, but at a further distance, would have to be so high that they exceed the finite escape velocity of our Galaxy \citep{koppelman2021determination} and to have many of these stars in the sample would be highly unlikely. The election of high tangential velocity (halo) main sequence stars in this RPM diagram allows us to create a sample with well-understood distances since main sequence stars have an approximately linear relationship between absolute magnitude and colour. 

In summary, the RPM diagram can be constructed using a combination of Gaia observables that are reliably available for a very large sample of its stars (apparent magnitude and proper motions), as shown in Eqn. (\ref{eq:rpm1}). It then returns valuable information on the absolute magnitude -- and thereby distance and tangential velocity -- of the stars in selected areas of the RPM space where main sequence halo stars live. 

\subsection{Photometric Distance Calibration}\label{2.2}
Most of the established methods to select halo stars involve the use of distance and spectroscopy information. By selecting the halo stars using the reduced proper motion method, we end up having a halo catalogue that is independent of using distances as an input. In turn, we aim to calculate the distances to these stars by using the apparent Gaia magnitude in the G-band. This colour absolute magnitude relation is simply referred to as photometric parallax in literature \citep{juric2008milky}. Photometric distance in kpc as a function of Gaia photometry and colour is given by the following equation:
\begin{equation}
    d_\text{phot}=10^{\frac{m_\text{G,0}-M_\text{G,0}-10}{5}}
	\label{eq:dist1}
\end{equation}
The errors in absolute magnitude are computed as the width of the absolute magnitude range for all the stars with high tangential velocities and good parallax in the RPM diagram in each of the colour bins. The percentage error in $d$ after propagating the errors in absolute magnitude (with the assumption that the errors in apparent magnitude are negligible) is given by
\begin{equation}
    \frac{{\delta}d_\text{phot}}{d_\text{phot}} {\times} 100=20\log(10){\delta} M_\text{G,0}
	\label{eq:dist2}
\end{equation}
where we assume an approximately linear relation between colour and absolute magnitude, which is a good approximation for the main sequence. In this work, we choose to use $G_0-G_\text{RP,0}$ colour, as $G_\text{BP,0}$ flux tends to be biased and overestimated for fainter sources. Using it would therefore be less ideal as it needs to be filtered creating selection effects \citep{riello2021gaia}. 

\subsection{\emph{Gaia DR3} halo cartograph}\label{2.3}

\subsubsection{Extinction correction and quality cuts} \label{subsubsection:ext-quality-cuts}

For the reduced proper motion halo selection using the recent Gaia DR3 catalogue we perform several photometric as well as astrometric quality cuts, and we remove sources with high extinction. The photometry used for analyses in this paper is corrected for interstellar extinction using \citet{article} 2D dust maps and following the procedure of \citetalias{Koppelman_2021} with small modifications. We will briefly outline the steps below.

The 2D dust maps account for the entire ISM dust along the line-of-sight. Following \citet{binney2014new} who use the relation by \citet{Sharma} for the dust density model, we are able to calculate a parameter called the ``extinction fraction'' that accounts for only the amount of foreground dust for each source based on its location in the sky. This fraction particularly makes a difference in extinction correction for sources that are within the solar neighbourhood. A simple analytical form of how we calculate the extinction fraction called the $N_{\text{ext}}$ (short for normalised extinction) in V-band is,

\begin{equation}
    N_{\text{ext}}=\frac{A_V(l,b,s)}{A_{V,\infty}(l,b)}=\frac{\int_{0}^{s}\rho(\Vec{x}(s^{'}))ds^{'}}{\int_{0}^{\infty}\rho(\Vec{x}(s^{'}))ds^{'}}
	\label{eq:ext-frac}
\end{equation}

where the denominator is the extinction value given directly from the 2D dust maps and the numerator is the amount of extinction for a star that is at a distance $s$ in kpc from the sun with the direction vector $\Vec{x}$ along the latitude and longitude $(\ell,b)$ in radians and the function represented by $\rho(\Vec{x}(s^{'}))$ inside the integral is the dust density which is adopted from Eqn. (16) in \citet{Sharma}. For sources with parallax < 0.1 mas (i.e., distant stars), we fix the $N_{\text{ext}}$ to be 1.0 to make the process computationally feasible and faster.    

We also scale the 2D dust maps in regions where $E(B-V)>0.15$ because they are overestimated \citep{arce1999measuring}. For this, we use the equation from \citet{binney2014new}. The scaling factor is estimated as follows:

\begin{equation}
    E(B-V)_\text{correction}=0.6+0.2\Bigg\{1-\tanh{\Bigg( \frac{E(B-V)-0.15}{0.3}\Bigg)}\Bigg\}
	\label{eq:ebv-correction}
\end{equation}

It is noteworthy that these corrections don't have a large impact on the selection of halo stars for the purposes of this study. However, the extinction correction process that we follow here can have an effect on sources in the solar neighborhood as well as more highly reddened regions. 

We use an extinction curve with $R_V=3.1$ that is computed using Padova model\footnote{\url{http://stev.oapd.inaf.it/cgi-bin/cmd_3.7}} which is originally based on \citet{extinction1} and \citet{extinction2}. Using this tool for Gaia DR3 photometric system, we apply $\frac{A_G}{A_V}=0.83627$, $\frac{A_{BP}}{A_V}=1.08337$ and $\frac{A_{RP}}{A_V}=0.63439$ to obtain the extinction correction for each Gaia passband.
Based on Section 8.3 from \citet{excess_flux}, we solve for a correction to the internally calibrated mean source G-band photometry to account for the systematic effect due to the use of default colour in the Image Parameter Detection (IPD). The correction is represented by a simple cubic polynomial as a function of BP-RP colour for different ranges of G-band, the coefficients of which are taken from Table 5 in \citet{excess_flux}. The corrected G-flux is a product of the given G-flux and this correction factor while the corrected apparent G-magnitude is $m_\text{G}-2.5\times\log(\texttt{correction-factor})$.  We calculate a correction for \texttt{phot\_bp\_rp\_excess\_factor} using values based on Table 2 in \citet{excess_flux} which gives us the value of what we call \texttt{excess\_flux}. We also calculate one sigma deviation for the \texttt{excess\_flux} for a sample of sources with good quality Gaia photometry using equation (18) from \citet{excess_flux}. Using these two parameters, we place a photometric quality cut on the full Gaia DR3 sample that is defined as $abs(\texttt{excess\_flux})<5\times\texttt{sigma\_excess\_flux}$. Additionally, we filter the sources with bad astrometric solutions by removing all the sources with \texttt{ruwe} > 1.4 (See more information on this in \citealt{Parallax} and \citealt{ruwe}). We remove high extinction sources using the quality cut $A_v<2.0$, which also acts as a way to remove disc contamination as most of the high extinction stars come from lower latitudes close to the disc. The mean values of $E(B-V)$ and the extinction normalisation factor in the V-band $N_{\text{ext}}$ in the final catalogue are 0.19 and 0.92 respectively.  

\subsubsection{Fitting the main sequence sources}\label{2.3.2}
Producing a reliable fit for the absolute magnitude of main sequence stars for different values of Gaia colour is one of the important steps in generating photometric distances to the sources in this halo catalogue. Instead of relying on isochrones, which are known to not always describe the data well in this colour-ranges \citep{blue-red-sequence}, we use an empirical relation based on real data. We select stars with high tangential velocities ($v_\text{tan}>200$ \text{km/s}) that also have good quality parallaxes (\texttt{parallax$\_$over$\_$error}$>5$) in the Gaia colour range $0.35<G_0-G_\text{RP,0}<1.1$ and impose a three component linear fit in absolute G-magnitude ranges between 4 and 6, 5 and 8, and above 8. After applying a zero point offset of 17 $\mu \text{as}$ to the Gaia DR3 parallaxes \citep{Parallax}, we invert them to use an estimate of distances to get the absolute magnitudes for this fit. We also perform a more precise running mean and standard deviation fit for the absolute magnitude by binning the colours into 128 components. We use the $\text{[M/H]} = -0.5$, 11 Gyr isochrone (obtained from \citealt{parsec-isochrone}) shifted by 0.01 mag in $G_0-G_\text{RP,0}$ to describe the valley between blue and red sequence. The blue sequence is inherently described as the halo stars with almost no rotation while the red sequence is heavily populated by heated-up thick disc stars with slow rotation \citep{koppelman2018one,blue-red-sequence}. All the stars to the right of this shifted isochrone are removed as red sequence contamination. This contamination is further reduced by using a stricter $v_\text{tan}>300$ km/s cut for the fitting purpose. We also remove residual stars below the main sequence using the following empirical cuts: $G_0-G_\text{RP,0}<0.65$ and $M_\text{G,0}>8$ or $G_0-G_\text{RP,0}<0.8$ and $M_\text{G,0}>10$.

\subsubsection{Selecting the final catalogue}\label{2.3.3}
We go back to the RPM diagram for the entire Gaia DR3 sample, now fully cleaned, and place a tangential velocity cut between 200 and 800 km/s in the main-sequence colour range. We then assign photometric distances to each of these stars based on the 3-component linear and running mean main sequence fit. The corresponding extinction-corrected RPM diagram for the entire Gaia DR3 data that passes the photometry, astrometry and extinction cuts as described in subsubsection \ref{subsubsection:ext-quality-cuts} is shown in Fig. \ref{fig:rpm}. The grey polygon is drawn based on the 3-component fit to the main sequence at 200 and 800 km/s tangential velocities. The red and blue isochrones (converted to RPM sequence using equation (\ref{eq:rpm2})) are taken from \citet{parsec-isochrone} with $\text{[M/H]} = -1.6$ and 12 Gyr age corresponding to the average metallicity and age of the local halo \citep[e.g.,][]{youakim2020pristine} to illustrate that the high $v_\text{tan}$ cut we impose on the main sequence picks up halo stars indeed. 

White dwarfs are removed by excluding all the sources below the 3-component fit on the main sequence offset by 2 mag in $M_\text{G,0}$. Because the uncertainties in proper motions and photometry are not large enough for the intrinsically brightest giants on the higher Red Giant Branch (hereafter RGB) to be picked up in our sample, we claim that the giant contamination is negligible. As a final quality cut, we place an empirical cut on reduced proper motion uncertainty in order to remove contamination from the lower RGB stars bleeding into the main sequence selection we use. The quality cut we impose on Gaia DR3 reduced proper motion parameter over reduced proper motion uncertainty (computed by propagating the errors in apparent G-magnitude and proper motions) is to keep all the sources that satisfy the following condition: $\log\frac{H_G}{\delta H_G}>1.75$.

\subsubsection{Final catalogue}\label{2.3.4}
The final catalogue comprises 47,650,376 provisional halo main sequence sources from Gaia DR3 selected using the reduced proper motion property. This is approximately five times larger than the number of sources presented in \citetalias{Koppelman_2021} using Gaia DR2. A sub-sample of this final catalogue with reliable photometric distances that excludes turnoff and redder stars by considering only stars with $0.45<(G_0-G_\text{RP,0})<0.715$ has a total of 24,647,379 stars, which is 3.5 times bigger than the one produced with Gaia DR2. Fig. \ref{fig:full-sky-sample} shows the on-sky density distribution of tentative halo main sequence stars from Gaia DR2 \citepalias{Koppelman_2021} and Gaia DR3 (this work) for different distance bins (All, $d>5$ kpc and $d>10$ kpc) in Galactic coordinates colour-coded by the logarithm of the number of stars present in each pixel produced using a k=12 HEALPix pixel level. The bottom middle panel shows the distribution of sources that have a heliocentric distance of more than 5 kpc which equals to $\sim75{\%}$ of the total sample while the bottom right panel shows the distribution of sources that have a heliocentric distance of more than 10 kpc that makes up $15{\%}$ of the total sample. The mean distance of the Gaia DR3 RPM selected halo sample is $\sim$6.6 kpc and goes out to 21 kpc which is much farther out than was possible using the catalogue derived for Gaia DR2 whose mean distance was 4.3 kpc. This is further illustrated in Fig. \ref{fig:full-sky-sample}. The on-sky distribution of the new catalogue looks more complete at distances greater than 5 kpc. Pixels with low/no stars in all three panels (especially at higher distances) correspond to high-extinction regions according to the Gaia dust maps. The Gaia scanning pattern created as Gaia has been scanning some regions more frequently is prominently visible in all the panels, because we probe the fainter stars pushing the limits up to Gaia's limiting apparent magnitude of 22 (see \citealt{excess_flux} for a clear view of Gaia's scanning pattern at fainter magnitudes). 

The resulting catalogue is available with this work. The first nine rows of the catalogue and their available parameters are shown in Table \ref{tab:mem}.

\begin{table*}
	\centering
	\caption{The first nine candidates and two stream member candidates in the reduced proper motion selected Galactic halo catalogue using Gaia DR3}
	\label{tab:mem}
	\setlength{\tabcolsep}{0.5em}
	\begin{tabular}{llllllllllllll} 
		\hline
	    Gaia source ID & RA & DEC & $\text{H}_\text{G}$ & $\text{A}_\text{V}$ & $\text{N}_{\text{ext}}$ & ${\text{d}_{\text{phot}}}$ & ${\delta \text{d}_{\text{phot}}}$ & Stream & ${\text{d}_{\text{[Fe/H],phot}}^{\text{}}}$ & ${ \text{d}_{\text{[Fe/H],phot}}^{\text{upper}}}$ & ${ \text{d}_{\text{[Fe/H],phot}}^{\text{lower}}}$ & $\text{v}_{\text{los}}$ & $\delta \text{v}_{\text{los}}$\\
	    & (deg) & (deg) & (mag) &(mag)  &   & (kpc) & (kpc) & & (kpc)&(kpc) & (kpc) & (km/s) & (km/s)\\
		\hline
        515396233856&	44.99&	0.06&	14.02&	0.23&	1.00&	5.8766&	0.4772\\
        5016521963392&	44.94&	0.13&	14.73&	0.21&	1.00&	4.4037&	0.2808\\
        6012955592192&	44.86&	0.17&	16.11&	0.21&	1.00&	1.5137&	0.2202\\
        8933531981952&	45.19&	0.17&	14.45&	0.26&	1.00&	5.5018&	0.7304\\
        9109626472192&	45.16&	0.18&	14.61&	0.27&	1.00&	1.9619&	0.1469\\
        10479720241408&	45.18&	0.23&	17.81&	0.27&	0.99&	1.4254&	0.2821\\
        12678743467648&	45.27&	0.31&	13.40&	0.28&	1.00&	5.2553&	0.8033\\
        12678743467776&	45.27&	0.31&	13.60&	0.28&	1.00&	4.3865&	0.3038\\
        13331578663424&	45.11&	0.24&	14.26&	0.27&	1.00&	5.0789&	0.4123\\
        ... &	... &	... &	... &	... &	... &	...	& ... & ... &	... & ... &	...	& ... \\
        1662967906302097536	& 201.37&	59.26&	14.27&	0.03&	1.00&	8.1565&	0.6125&	GD-1 & 7.5405 & 2.1557 & 0.6793\\
        832183620503381504 &	162.88 &	47.53 &	15.42 &	0.05 &	0.99 &	6.9873	& 0.4802 & GD-1 &	6.7155 & 1.4698 & 0.7509 &	-126.64	& 14.48 \\
		\hline
	\end{tabular}
\\Note: A full electronic version of the catalogue will be made available online as one full catalogue and each stream member as a separate catalogue. 
\end{table*}

\begin{figure}
    \centering
	\includegraphics[width=\columnwidth]{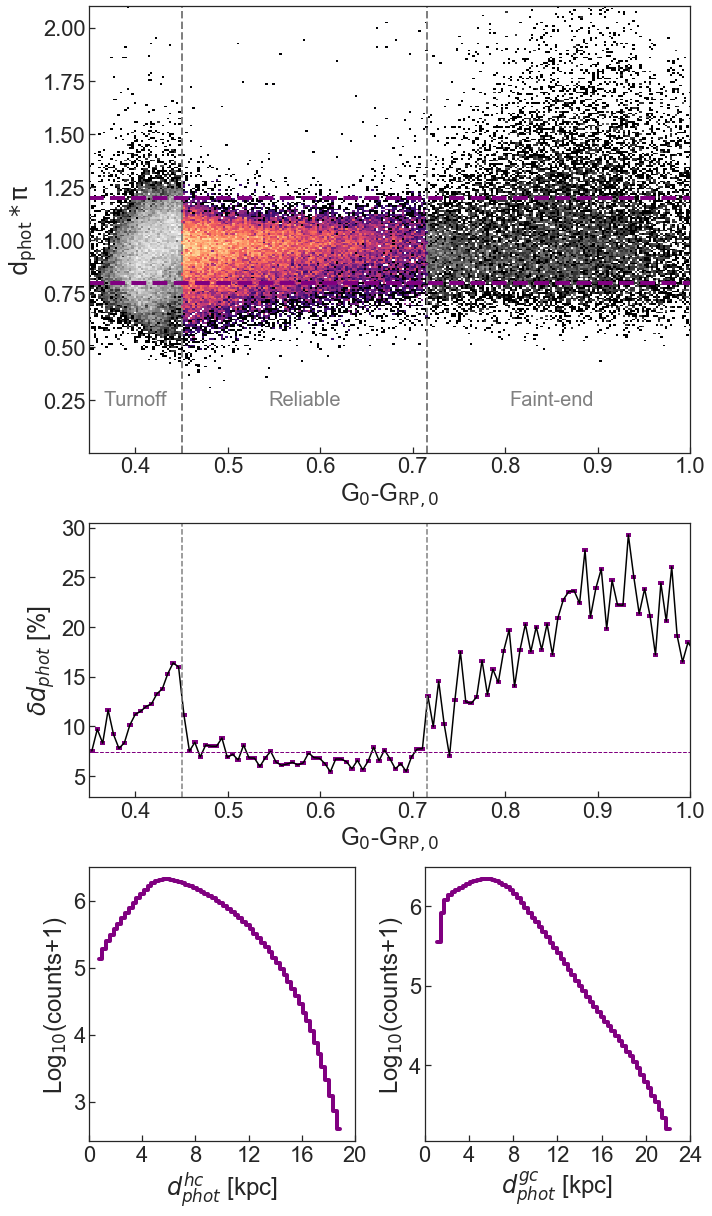}
    \caption{Quality and distribution of the photometric distances. Top panel: Comparison of photometric distance derived in this work to the Gaia good quality parallaxes as a function of Gaia colour. Purple thick dashed lines indicate ${\pm}10{\%}$ difference between photometric distances in this work and Gaia parallaxes. The colour range between which the sample produces reliable distances is marked with the label 'Reliable' and the subsample outside this range is marked as 'Turnoff' and 'Faint-end' respectively. Middle panel: Percentage error in photometric distances as a function of Gaia colour. Purple thin dashed line indicate the typical photometric distance uncertainty ($\sim7\%$) within the reliable Gaia colour range. Bottom panel: Distribution of heliocentric photometric distances (left) and galactocentric photometric distances (right) for the entire sample as derived in this work. Note that the y-axis is the logarithmic density in each bin.}
    \label{fig:distance-error}
\end{figure}

\begin{figure}
	\includegraphics[width=\columnwidth]{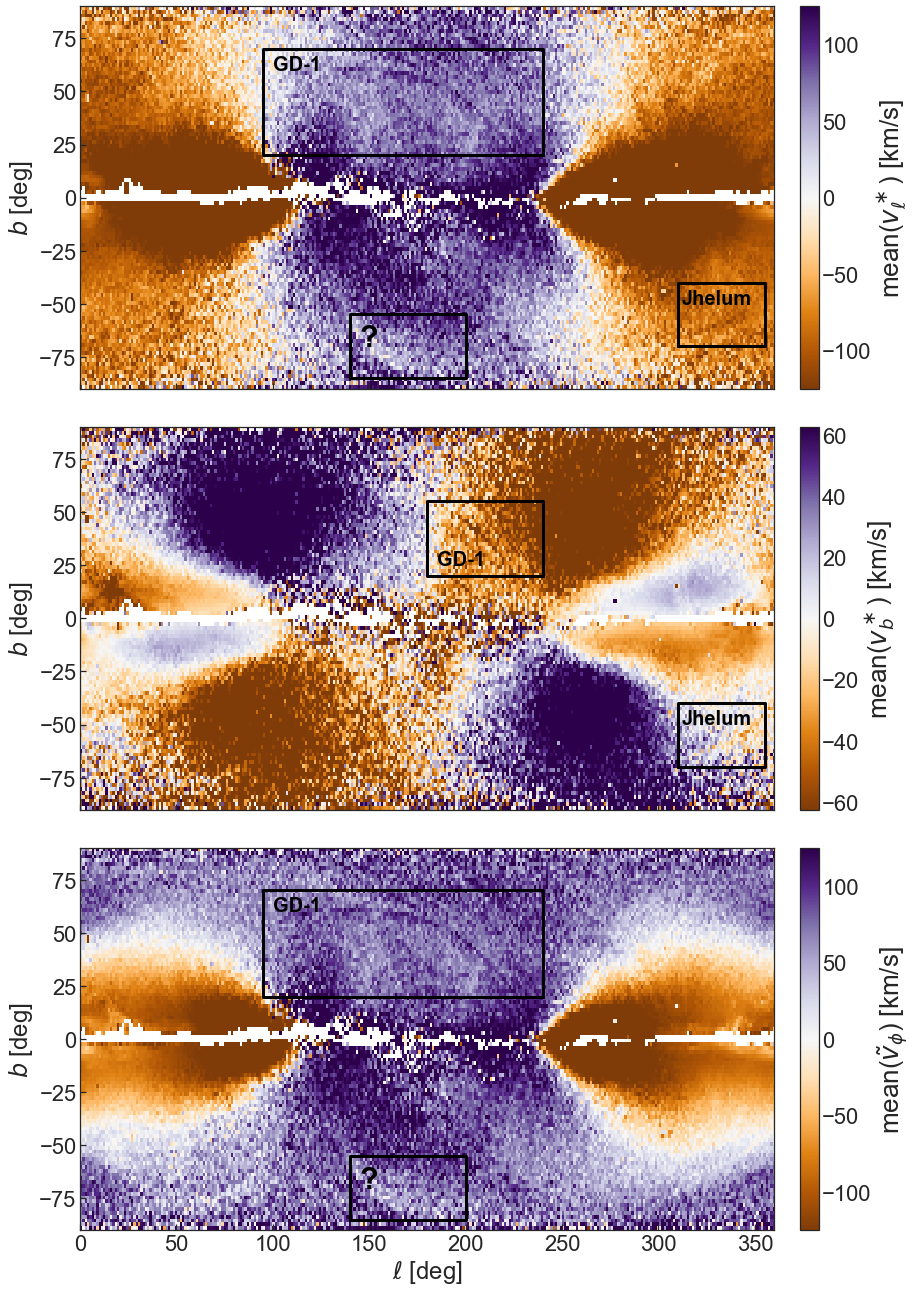}
    \caption{Galactic star map binned with the mean value of solar motion corrected longitudinal (top), latitudinal (middle), and pseudo azimuth velocities (bottom) in each pixel for the heliocentric distances above 7 kpc. Several coherent retrograde streams and substructures pop up in each of the three panels.}
    \label{fig:binned-velocity-sky}
\end{figure}

\subsection{Photometric distance validation}\label{2.4}
One of the major advantages of using this sample to study the halo is that we can derive reliable photometric distances to the main sequence halo stars. Therefore, it is important that we analyse the quality of the photometric distances we calculate. The textbook sample to compare with are the inverted parallaxes from Gaia for sources with good quality parallaxes, \texttt{parallax\_over\_error} > 20 within the RPM selected halo sample. We end up with $\sim$80,000 sources with good parallaxes which is less than $1\%$ of the entire sample size as Gaia parallaxes are unreliable for distant and faint sources. We use these 80,000 sources as a representative sub-sample to validate the quality of the photometric distances computed using the main sequence fit. The product of the photometric distance and parallax (which should ideally be 1.0) and the percentage error in the photometric distances that are calculated by propagating the uncertainties in the absolute magnitude calculated based on the running mean fit on each of the 128 Gaia colour bins (see equation (\ref{eq:dist2})) versus Gaia colour are shown in the top and middle panel of Fig. \ref{fig:distance-error} respectively. It is clearly evident from these figures that the photometric distances are more reliable, relatively speaking, inside the colour range: $0.45<(G_0-G_\text{RP,0})<0.715$. The typical uncertainty within this colour range is down to $7.4\%$ The sample within this colour range has a median distance error of $0.39_{-0.12}^{+0.18}$ kpc and a median velocity error taking into account the error in the distance as well as that in the proper motions of 15.5 km/s. The whole sample has a median distance error of $0.57_{-0.26}^{+0.56}$ kpc.  

Stars bluer and redder than this have typically larger relative uncertainties. On the one hand, the typical uncertainty in the bluer main sequence turnoff part with $(G_0-G_\text{RP,0})<0.45$ is $12\%$. This larger parallax-distance deviation near the main-sequence turnoff is caused by the almost vertical nature of this part of the HR diagram; this makes absolute G-magnitude as a function of colour more uncertain. Typically, our method tends to underestimate the distances for these stars which explains the overpopulation of stars at the MSTO colour range below 1.0 in the y-axis in the top panel of Fig. \ref{fig:distance-error}. One illustration of the possible consequences of this underestimation is shown in Section \ref{3.4}. On the other hand, for redder sources than $G_0-G_\text{RP,0}>0.715$ the typical distance uncertainty inflates to $16\%$ and we see from the parallax-distance comparison that the distances are typically overestimated (see the stars at the redder range above 1.0 in y-axis in the top panel of Fig. \ref{fig:distance-error}). The cause of this is a broadening of the main sequence at the faint end due to a lack of faint stars (Gaia's incompleteness from G > 20) increasing the uncertainty in the calibration of photometric distances in this range. 

The bottom panel of Fig. \ref{fig:distance-error} shows the distribution of heliocentric and galactocentric distances. The two local peaks in the galactocentric distance distribution are attributed to the overpopulation of stars around the solar neighbourhood as expected. Thus, the distances agree well with the parallaxes in the reliable colour range and the distribution agrees with the hypothesis that the Milky Way stellar halo has a steep density profile as presented in the literature so far \citep{deason2011milky}.

\section{Substructure(s) in the binned velocity space}\label{3}

Hunting for spatially coherent halo substructures in the sky is possible using proper motion information while the structures are hidden in a maze of smooth background halo distribution. An even more reliable approach is to combine proper motion with the distance information and plot the mean tangential velocity components of each pixel in the sky to look for cold streams and/or over-densities. This method will work only if the velocity vectors in the sky are sufficiently different from the background velocity distribution and it is also sensitive to the velocity dispersion of the structure itself and uncertainties on the velocities. In this section we will use this approach on the RPM sample. It is important to remember here though that: (i) the RPM selected halo sample suffers from a kinematic selection bias and (ii) the distances are not very accurate close to the turnoff and for fainter stars with redder colours. 

To find the on-sky velocity components, we use the following equation:

\begin{equation}
    v_\text{i}=4.74057\times{\mu_\text{i}}\times{d_\text{phot}}
    \label{eq:v1}
\end{equation}

where i can be the Galactic longitude or latitude in radians to derive their respective velocity components in km/s. It is important that these space velocities are corrected for solar reflex motion using the distances that we calculate. For this purpose, we calculate the solar velocities at each (l,b) using the following equation:

\begin{gather}
    v_{{l},\odot}=-U_\odot\sin{l}+(V_\odot+V_\text{LSR})\cos{l}\\
    v_{{b},\odot}=W_\odot\cos{b}-\sin{b}(U_\odot+(V_\odot+V_\text{LSR})\sin{l})
    \label{eq:v2}
\end{gather}

where we use the solar motion constants $(U_\odot,V_\odot,W_\odot)=(11.1,12.24,7.25)$ km/s and the local standard of rest motion $V_\text{LSR}=232.8$ km/s from \citet{schonrich2010local} and \citet{mcmillan2016mass} respectively. Now we add the solar correction to the velocities we calculated in equation (\ref{eq:v1}) to get the solar motion corrected Galactic space velocities in km/s.

\begin{equation}
    \label{eq:v5}
    v_\text{i}^*=v_\text{i}+v_{\text{i},\odot}
\end{equation}

where $i$ can be $(l,b)$ in radians. 
In this halo main sequence catalogue, the mean uncertainty in velocities for the entire sample within the $G_0-G_\text{RP,0}$ colour range where the distances are more reliable are $\overline{v_l^*}\sim23.6$ km/s and $\overline{v_b^*}\sim17.9$ km/s. The mean errors for turnoff and fainter stars are comparatively larger: $\overline{v_l^*}\sim37.2$ km/s and $\overline{v_b^*}\sim24.8$ km/s.

Only 12,862 stars in our catalogue have radial velocity measurements from Gaia because of the intrinsically faint tracers of the sample. Cross-matches with spectroscopic surveys such as LAMOST, SDSS SEGUE, APOGEE and GALAH give us less than 100,000 stars in common altogether. As this would severely restrict our search, we compute pseudo-3D-velocities instead by assuming the radial velocity component to be zero i.e., ${v_\text{los}^*}=0$. Taking into account the local standard of rest, the motion of stars around the Galactic Centre follows a $\sin{l}\cos{b}$ pattern which can be put to use without line-of-sight velocities in certain regions in the sky, such as near both Galactic poles, near Galactic centre and anti-centre \citep{Kim_2021}. However, here we are trying to look for streams and substructures on the entire sky. Considering that, we choose to compute pseudo-velocities for the entire sample in $(x,y,x)$ and $(R,\phi,z)$ coordinates. The pseudo velocities in km/s in 3D cartesian coordinates are given by,

\begin{gather}
    \label{eq:v4}
    \tilde{v}_x=-v_l^* \sin{l} - v_b^* \cos{l} \sin{b}\\
    \tilde{v}_y=-v_l^* \cos{l} - v_b^* \sin{l} \sin{b}\\
    \tilde{v}_z=-v_b^* \cos{b}
\end{gather}
The tilde symbol above the velocity representation is to indicate that these are not true 3D velocities. We can perform coordinate transformation on these values to get pseudo-3D-cylindrical-velocities namely $(\tilde{v}_R,\tilde{v}_{\phi},\tilde{v}_z)$. All these pseudo velocities are in galactocentric coordinates by placing the sun at $(X_{\odot},Y_{\odot},Z_{\odot})=(-8.2,0.0,+0.014)$ kpc \citep{GRAVITY2018}. 

The sky distribution in Galactic coordinates $(l,b)$ showing the mean velocities of stars in pixels of $360/300 \times 180/100$ is shown in Fig. \ref{fig:binned-velocity-sky}. We see several members of streams that move in the same direction with a velocity that is sufficiently different from the nearby background. The top panel is binned for velocity distribution in the longitude direction, the middle panel is binned for velocity distribution in the latitude direction and the bottom panel is binned for velocity distribution in the pseudo azimuth direction. All these sub-figures have candidate stars selected with a photometric distance greater than 7 kpc (distant halo main sequence stars). We are mainly concerned about the velocity distribution at (comparatively) higher distances because streams, substructures and over-densities that are in the solar neighbourhood will not typically appear as cold and coherent structures in the sky. We choose a distance greater than 7 kpc because it is the mean distance of the sample. It is also important to note that we plot all the distant halo main sequence stars in our sample including the turnoff stars for which the uncertainties in the photometric distances can be relatively high. 

In the following subsections, we study the two most obviously visible structures in the sky to the faintest Gaia G magnitudes and characterise them independent of any prior information about these structures from the literature. These selection methods, in principle, can be used for any streams picked up using the RPM sample. 

\subsection{GD-1 stellar stream}\label{3.2}
One of the most easily visible structures in the northern hemisphere in Fig. \ref{fig:binned-velocity-sky} is the very retrograde GD-1 stream discovered using Sloan Digital Sky Survey by \citet{Grillmair_2006}. In the past 1.5 decades, this stream has been studied in unprecedented detail (for some most recent studies, see \citealt{ibata2020detection,balbinot2021,banik2021evidence,dillamore2022impact,doke2022probability,shih2022via}).

\begin{figure}
    \centering
	\includegraphics[width=\columnwidth]{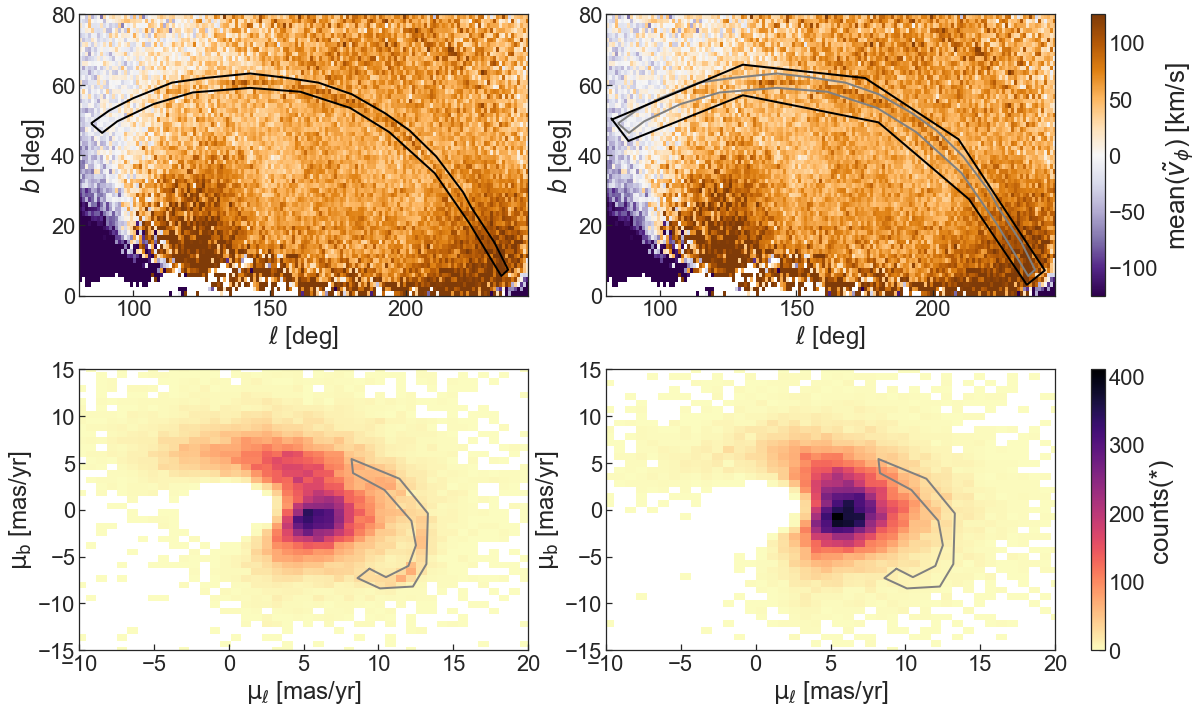}
    \caption{Binned pseudo azimuth velocity moments at d>6 kpc in the sky with the (rough) polygon selection of the GD-1 stream (top) and Gaia proper motions in (l,b) (bottom) within the polygon selection of the stream - on stream track (left) and outside the polygon selection of the stream - off stream track (right).}
    \label{fig:gd-1-selection}
\end{figure}

To pick up candidate stars that could potentially belong to this stream, we draw an empirical polygon around the pixels that distinctively vary in mean velocity along the longitude direction from the nearby background halo in the sky. We randomly select the same number of candidate stars in all directions around the stream structure in the nearby halo as the control sample. The sky space and the rough polygon selection in and around the stream are presented in the top panel of Fig. \ref{fig:gd-1-selection}, where the black polygon on the top left shows the GD-1 candidate members selected and the black polygon (after removing the stars in the grey polygon belonging to GD-1) on the top right shows the control sample of candidate members around the stream. When we plot the non-solar motion corrected Gaia proper motions in latitude and longitude direction of these on-stream and off-stream track members selected from the top panel, we clearly see an arc of proper motion peaks belonging to the GD-1 main sequence candidates to the right of the background halo proper motion peaks. The 2D proper motion plot and the selection of proper motion peaks belonging to GD-1 candidate stars are shown in the bottom panel of Fig. \ref{fig:gd-1-selection}. We can clearly see the kinematic selection effect (high tangential velocity selection) of the RPM sample as the void of halo stars with small proper motions (upper right part) in this subplot which is also the reason why we refrain from using complex statistical selection methods like Gaussian mixture models for picking up the proper motion peaks of the stream here and throughout the rest of this paper.

In order to validate the candidate members selected using proper motion peaks, we fit a third degree polynomial to the proper motion in right ascension and declination with respect to the stream coordinate $\phi_1$ using the stream coordinate system and conversion defined by \citet{koposov2010constraining}. Throughout the rest of this work, we only fit polynomial functions for the removal of contaminants and validation to account for the lack (or minimal availability) of line-of-sight information in these faintest magnitudes and to avoid biases and errors that may arise from using different Galactic potential models. In all the fitting polynomials, $\phi_1$ and $\phi_2$ are in units of radians. Because the stream selection we perform does not pick up members from the lower end of $\phi_1$, we place a cut on the lower limit of $\phi_1$ and remove obvious contaminants at the edge of the grid before we fit an empirical polynomial to these candidates. These polynomial functions match very well (with slight variations that we believe to be due to the incompleteness in lower $\phi_1$ ranges) with the ones defined by \citet{ibata2020detection}. We select the candidates that lie within $2\sigma$ standard deviation from the mean error ($\delta$) in proper motion. This is expressed by the following equation(s):

\begin{gather}
    \mu_\text{RA} \pm (\overline{\delta(\mu_\text{RA})}+2\sigma_{\delta(\mu_\text{RA})}) < 3.4\phi_1^3+8.25\phi_1^2+1.25\phi_1-7.5\\
    \mu_\text{Dec} \pm (\overline{\delta(\mu_\text{Dec})}+2\sigma_{\delta(\mu_\text{Dec})}) < -2.2\phi_1^3+5.2\phi_1^2+15.6\phi_1-4
\end{gather}

\begin{figure}
    \centering
	\includegraphics[width=\columnwidth]{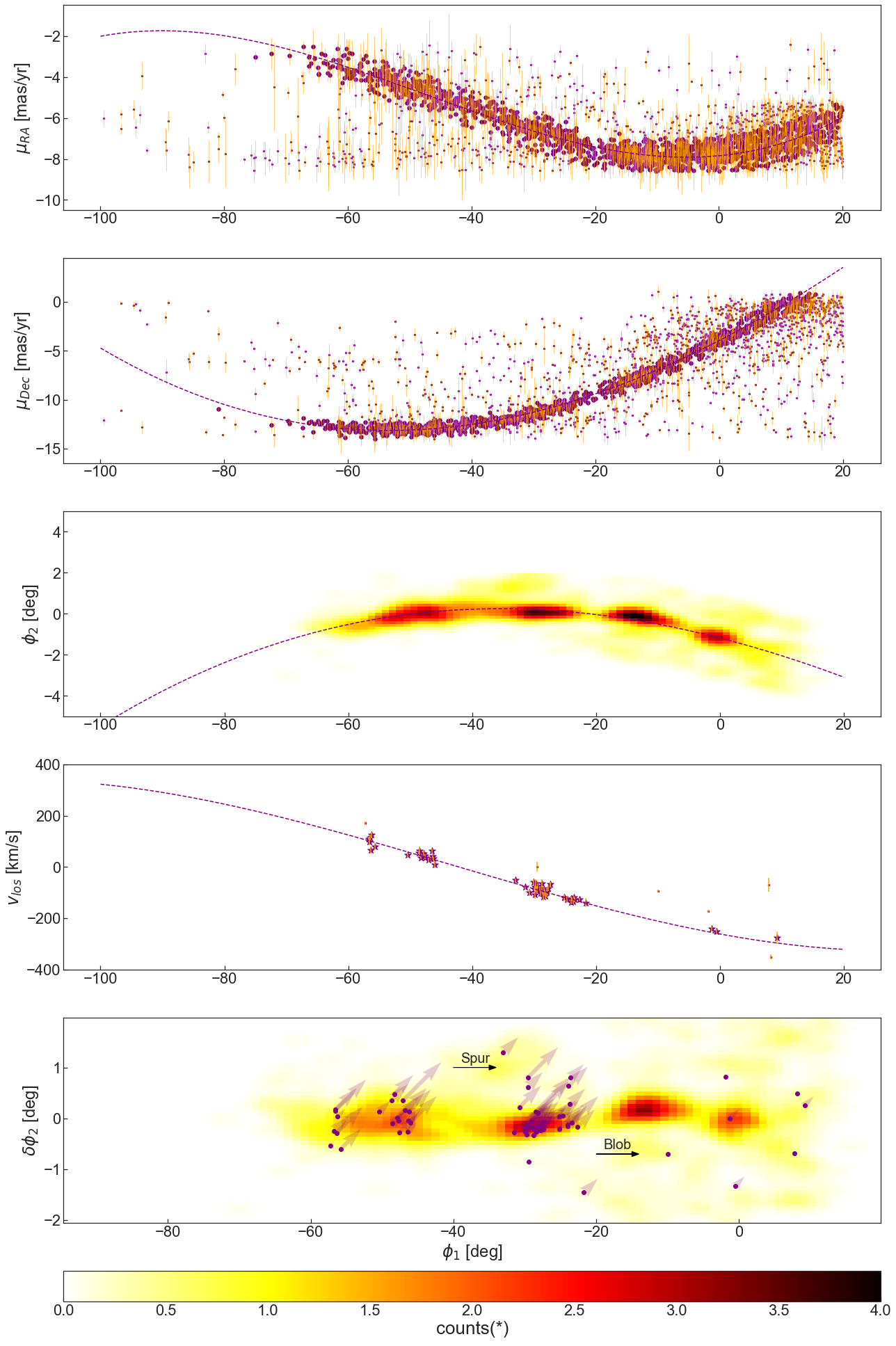}
    \caption{Proper motion in RA direction (top), Dec direction (middle top) with respect to the stream coordinate $\phi_1$, proper motion fitted confident members in GD-1 stream coordinates (middle), confident radial velocity members from SEGUE (middle bottom) and the stream track variations with respect to the coordinate $\phi_1$ with radial velocity members and confident candidates plotted as longitudinal velocity vector quivers (bottom). The dashed lines in plots one through four are third degree polynomial fits on the corresponding parameters. The counts(*) for the Gaussian smoothed density plot denote the number of stars per $1.2\times0.2$ $\text{deg}^2$ and $1.2\times0.08$ $\text{deg}^2$ in panel 3 and 5 respectively.}
    \label{fig:gd1-validation}
\end{figure}

where $\mu_\text{RA}$ and $\mu_\text{Dec}$ are in mas/yr. The selection polynomial and the method are illustrated in the top two panels of Fig. \ref{fig:gd1-validation}. All the stream star candidates that satisfy the conditions described by the above equations are chosen to be the confident members and plotted on the stream sky coordinates $(\phi_1,\phi_2)$. A third degree polynomial fit to the end stream track is described by the following equation:

\begin{equation}
    \phi_2 = F(\phi_1) = 0.00804\phi_1^3-0.055\phi_1^2-0.077\phi_1-0.021
\end{equation}

This equation also matches well with the polynomial stream track derived in \citet{ibata2020detection}. These confident members are shown in the middle panel of Fig. \ref{fig:gd1-validation}. To evaluate the performance of this independent stream candidate selection using the velocity space of the reduced proper motion selected halo sample, we cross-match the final members with publicly available existing spectroscopy surveys and find 64 stars with line-of-sight velocities and metallicities from SEGUE (Sloan Extension for Galactic Understanding and Exploration). The line-of-sight velocities for these stars from SEGUE versus stream coordinate $\phi_1$ are shown in panel 4 of Fig. \ref{fig:gd1-validation}. The empirical stream track is fitted using the polynomial described by equation (1) from \citet{ibata2020detection}. We select the confident radial velocity members using the following condition:

\begin{equation}
    v_\text{los} \pm (2{\delta v_\text{los}}+20) < 90.68\phi_1^3+204.5\phi_1^2-254.2\phi_1-261.5
\end{equation}

where $v_\text{los}$ is in km/s. Out of the 64 cross-matches, 58 (91$\%$) of them check to be confident members under the condition described by the above equation. The ones that do not satisfy this condition still look coherent in the position-velocity space. Confident members and radial velocity members (plotted as velocity quivers) are shown as smoothed density plots in $\phi_1$ versus $\delta\phi_2=\phi_2-F(\phi_1)$ in the bottom panel of Fig. \ref{fig:gd1-validation}. We are able to see the spur and diffuse blob feature at $(\phi_1,\phi_2)\sim(-30,1)$ and $(\phi_1,\phi_2)\sim(-20,-0.5)$ that was discovered by \citet{price2018off} which were proposed be caused by dark matter substructures in the Milky Way \citep{bonaca2020high}. The spur and blob are underdense by $\num{\sim3}\sigma$ significance compared to the highest density stream track component. We also see three gaps and/or density variations across the stream at $\phi_1\sim-38,-20,-3$ with $\num{\sim1}$, 2, and 3$\sigma$ significance. These gaps were also confirmed by \citet{de2018deeper} and \citet{de2020closer}. The member candidates away from the stream track are contaminated due to the edge selections in proper motion that can also be seen as crowding in top panels of Fig. \ref{fig:gd1-validation} at higher values of $\phi_1$. More information in 6D space and chemistry is needed to confirm their membership, or refute it. 

\begin{figure*}
    \centering
	\includegraphics[width=\textwidth]{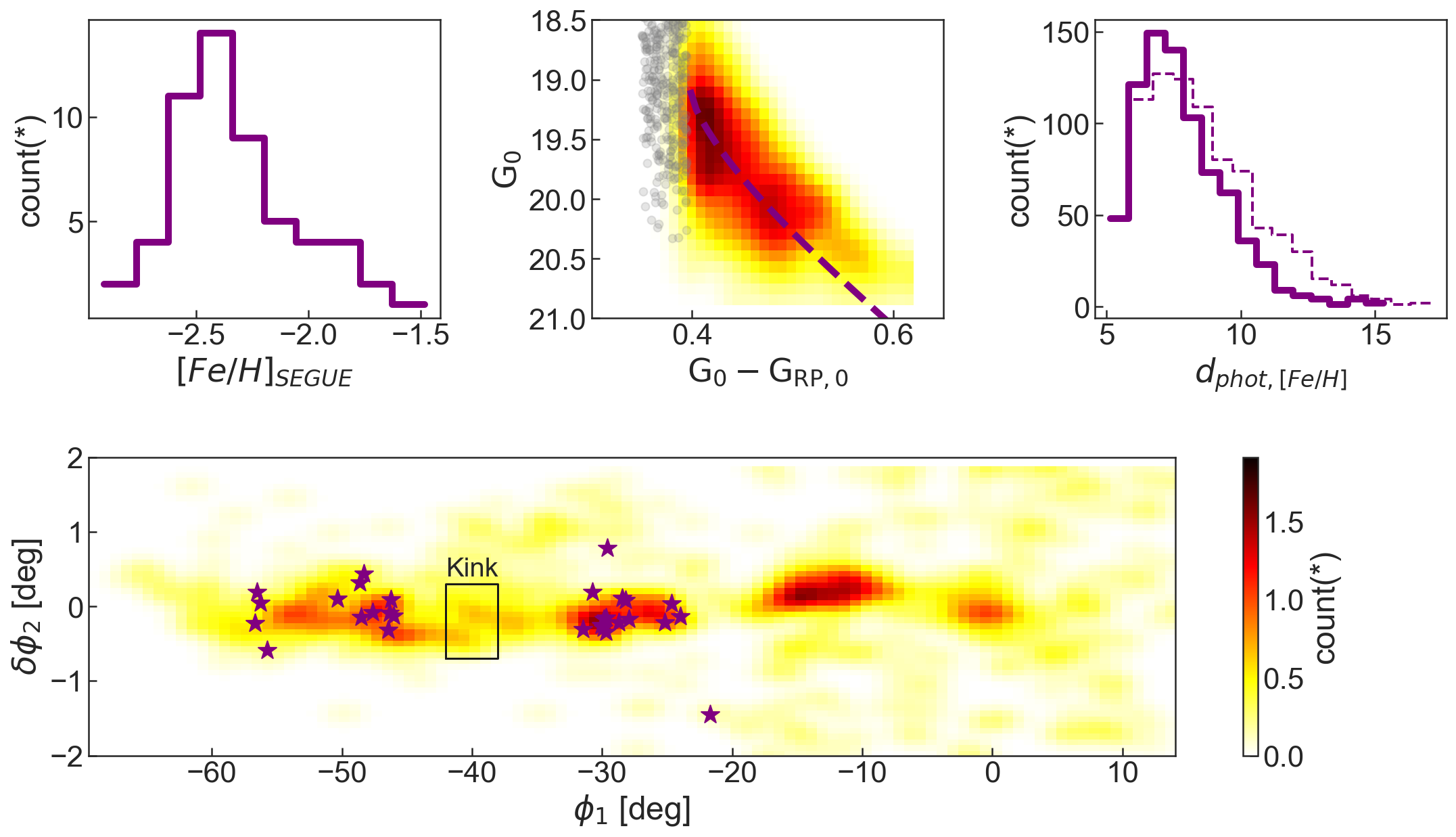}
    \caption{Metallicity distribution of radial velocity members (top left), colour-magnitude Diagram density distribution of confident main sequence with a -2.29 [Fe/H], 8.06 kpc (value taken from \citealt{malhan2022global}) mock isochrone in purple and non-main sequence in grey (top middle), metallicity sensitive photometric distance distribution as a continuous line and photometric distance distribution as a dashed line (top right) and the stream track variations with respect to the coordinate $\phi_1$ with radial velocity main sequence members (bottom). Note that this is a subset of the stars used for the bottom panel of Fig. \ref{fig:gd1-validation} where also turnoff stars were included.} The counts(*) for the Gaussian smoothed density plot is simply the number of stars per $0.8\times0.1$ $\text{deg}^2$. Note a discontinuity/kink clearly visible at $\phi_1\sim-40^{\circ}$.
    \label{fig:gd1-distance}
\end{figure*}

To further improve the analysis of the fainter counterparts of stream star candidates, we make the photometric distances more accurate by folding in the (known) metallicity distribution of the stream into distance calibration. Again, we refrain from using existing publicly available isochrones such as MIST \citep{choi2016mesa} and PARSEC \citep{parsec-isochrone} for this purpose, as our analysis shows the isochrones mismatch with each other in Gaia DR3 colours at these faint magnitudes on the main sequence and for $4<M_\text{G,O}<8$ and moreover deviate from local halo data with good parallax information, the latter of which is also suggested in \citet{Kim_2021} (hereafter \citetalias{Kim_2021})\footnote{After submission of this manuscript, \citet{2022MNRAS.515..795K} published more precise metallicity grids in an erratum. We have updated the published catalogue to make use of these. The change in distances is of the order of 1-2\% and the conclusions remain unchanged.}. This can be explained as the current stellar evolution tracks do not reproduce the cold low-mass stars very well. Instead, we choose to use the empirical photometric metallicity grids from Table 3 in \citetalias{Kim_2021} as mock isochrones to improve our photometric distances for the stream star candidates. \citetalias{Kim_2021} creates this photometric metalicity grid by selecting high tangential velocity local halo population out to $\sim$2 kpc using the RPM diagram, cross-matching them to existing spectroscopic surveys such as SDSS SEGUE I/II \citep{yanny2009segue}, SDSS APOGEE DR16 \citep{ahumada202016th}, LAMOST DR6 \citep{cui2012large}, GALAH DR3 \citep{buder2021galah+} and nearby main sequence from \citet{hejazi2020chemical} to get $\sim$20000 stars with metallicity information and building a photometric metallicity grid based on the distribution of the stars in the dataset. As this is a very similar selection to our sample, we deem these grids very justified for this work. One major difference between our sample and \citetalias{Kim_2021} sample is that it is not cleaned for thick disc contaminants, but as these will have metallicities significantly higher than the mean metallicity of GD-1 stream ($\overline{\text{[Fe/H]}}\sim-2.29$ based on SEGUE cross-match of candidates confirmed in this paper) this will not create any issues. 

Using these grids, we create a two dimensional interpolation grid based on metallicity and Gaia colour to give the absolute magnitude in G-band and thus photometric distances corrected for the metallicity of the stream. We find that the distance uncertainties are not affected in their magnitude by this change. The upper and lower sigma confidence intervals given in Table \ref{tab:mem} is therefore applicable to both sets of distances, provided that the shift between the general distance and metallicity-dependent distance measurements is taken into account. The final distribution of the distance is shown in the top right panel of Fig. \ref{fig:gd1-distance}. The distance re-calibrated with metallicity information and distance calculated for the entire RPM sample (without any prior information on metallicity) are shown as continuous and dashed line distribution respectively on the histogram. It is important to note that the distance histogram cannot be used at face value because of the incompleteness of the main sequence sample due to Gaia's incompleteness for sources with G < 19. We can clearly see from the CMD in the top middle panel of Fig. \ref{fig:gd1-distance} that we do not probe the entire magnitude range in every colour bin (as illustrated by the CMD, where we see that the range of magnitudes probed in the faint end of the Gaia colours - around $(G_0-G_\text{RP,0}) = 0.6$ - is much smaller than at $(G_0-G_\text{RP,0}) = 0.4$) and therefore, we are clearly missing a lot of distant main sequence in the sample due to Gaia's limiting magnitude. However, even an incomplete catalogue of faint main sequence stars belonging to the GD-1 stream is still an interesting probe into the low surface brightness components of this stream. We also show the sub-giant or turnoff stars, for which we have less accurate distances, in grey in the top middle panel of Fig. \ref{fig:gd1-distance}. These are still likely members of GD-1 but unlikely to be true main sequence stars (especially as the turnoff for such metal-poor systems is rather blue). 

The final catalogue of main sequence stars associated to GD-1 is plotted as smoothed density distribution in the bottom panel of Fig. \ref{fig:gd1-distance}. We are able to see a break in track (kink) at $\phi_1\sim-40$ as proposed by \citet{de2018deeper} but the shape of this kink is
opposite to the shape seen in their data while it matches the shape seen by the simulated stream from the same work. The two tracks are -0.12$^{\circ}$ and -0.06$^{\circ}$ away from the main stream track. However, their 16th and 84th quantiles overlap significantly, indicating that more spectroscopically confirmed members would be needed to make any statistically conclusive statements on these. \citet{webb2019searching} modelled the location of GD-1's progenitor using N-body simulations and data from Gaia DR2 and concluded that the stream’s progenitor could be located between $-30<\phi_1<-45$ and this could be responsible for the observed gap at $\phi_1\sim-40$. Following this argument, $\phi_1\sim-40$ could be a probable place for the GD-1 progenitor which we can see as a kink in our data. However, also here, more spectroscopic follow-up will be necessary to draw any firm conclusions. 

Some very distant $(|\delta\phi_2|>1)$ radial velocity members and small density peaks in our sample (see bottom panel of Fig. \ref{fig:gd1-validation} and \ref{fig:gd1-distance}) further illustrate the high complexity of this stream. 

\begin{figure}
    \centering
	\includegraphics[width=\columnwidth]{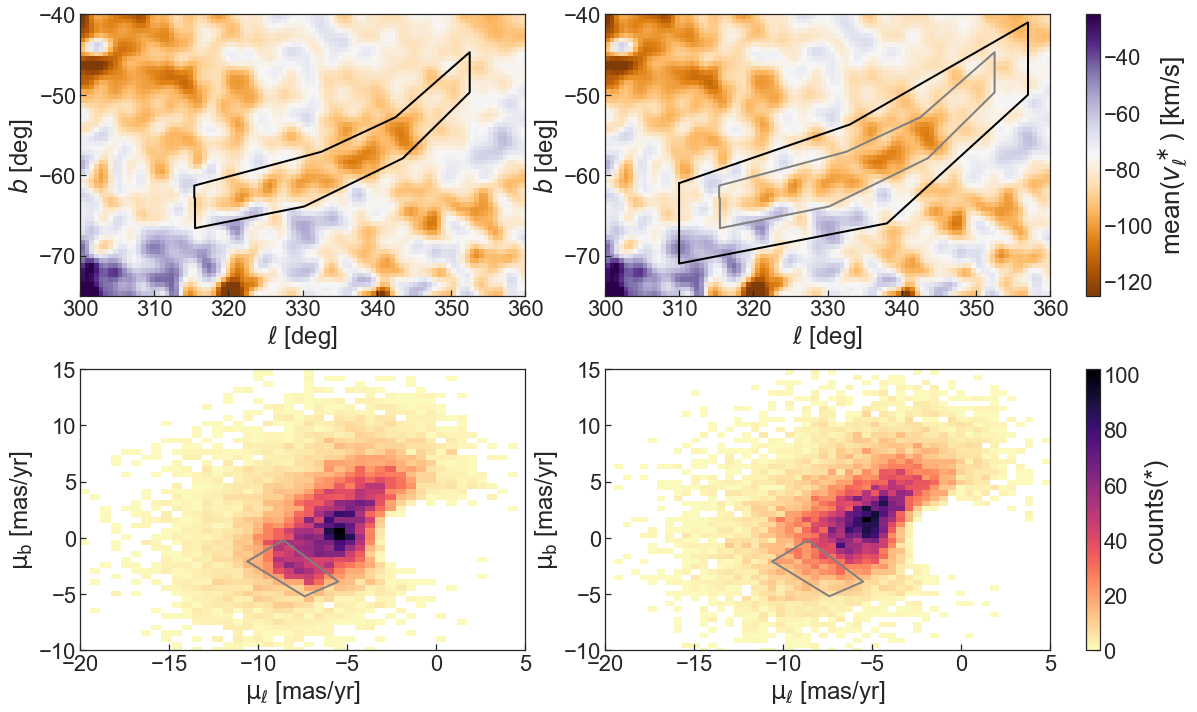}
    \caption{Binned longitudinal velocity moments in the sky at d>8 kpc with the (rough) polygon selection of the Jhelum stream (top) and Gaia proper motions in (l,b) (bottom) within the polygon selection of the stream - on stream track (left) and outside the polygon selection of the stream - off stream track (right).}
    \label{fig:jhelum-selection}
\end{figure}

\begin{figure}
    \centering
	\includegraphics[width=\columnwidth]{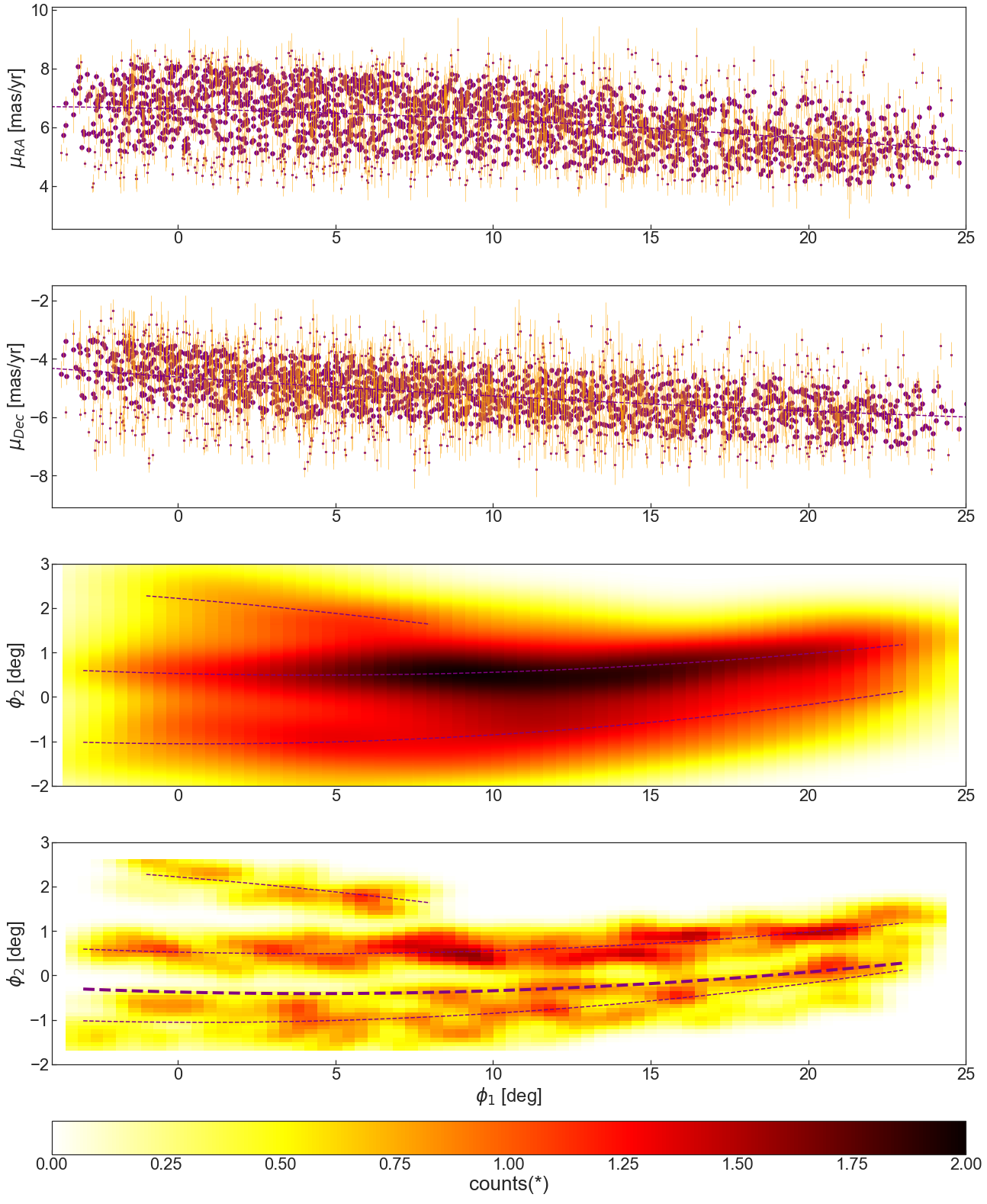}
    \caption{Proper motion in RA direction (top), Dec direction (middle top) with respect to the stream coordinate $\phi_1$, confident members on stream sky coordinates as KDE smoothed distribution (middle bottom) and the stream coordinates density distribution in the sky with the confident members of the three sub-components of Jhelum (bottom). The thin dashed lines in the plots are second degree polynomial fits on the corresponding parameters. The thick dashed line in the bottom panel refers to the equivalent fit by \citet{bonaca2019multiple} for the broad component (described as $0.9^{\circ}$ below the narrow component). The dashed lines in plots one and two are second degree polynomial fits on the corresponding parameters. The counts(*) for the Gaussian smoothed density plot is simply the number of stars per $0.4\times0.1$ $\text{deg}^2$.}
    \label{fig:jhelum-validation}
\end{figure}

\begin{figure*}
    \centering
    \includegraphics[width=\textwidth]{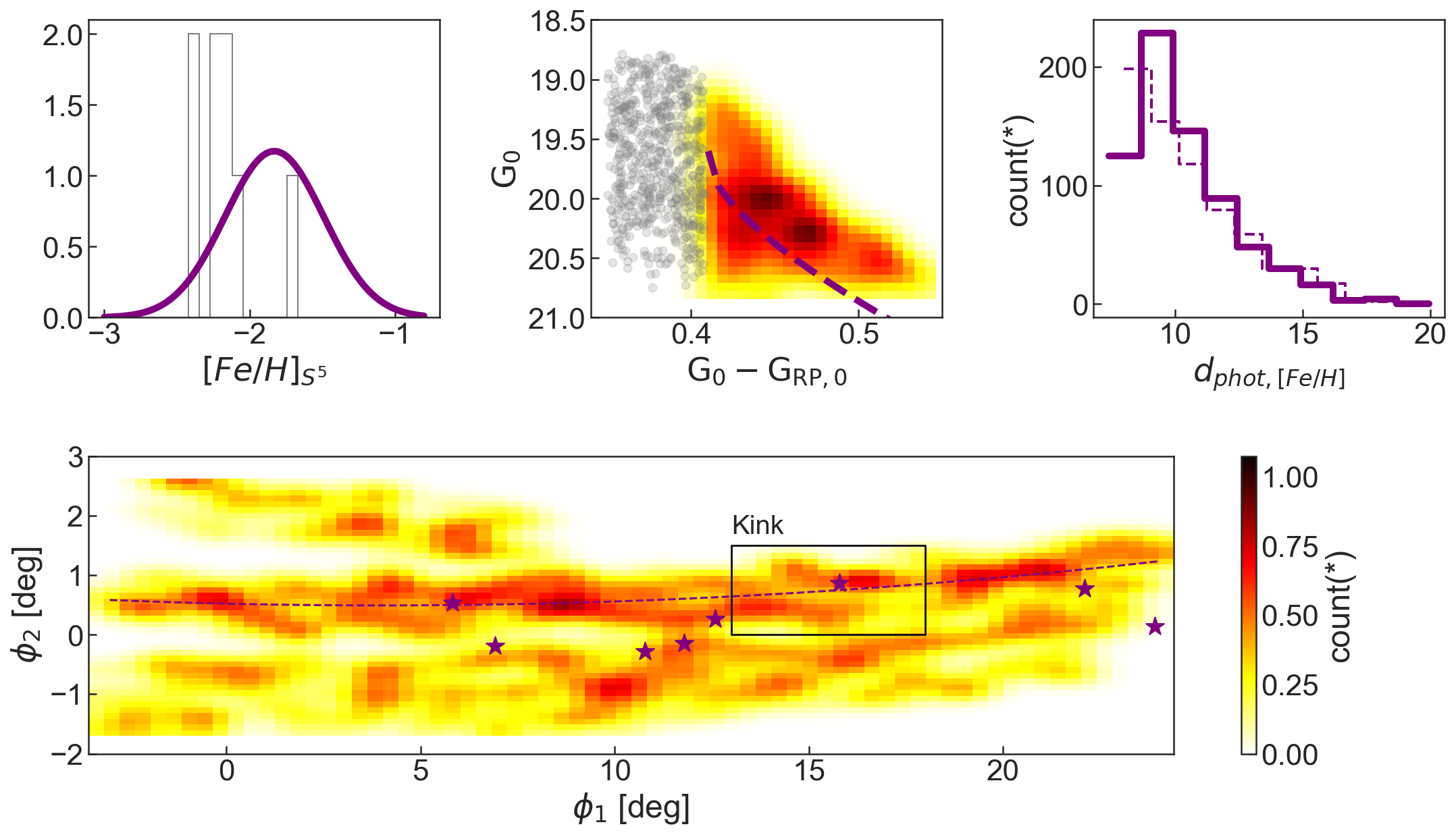}
    \caption{Assumed Gaussian distribution of metallicity based on \citet{li2022s} (top left), Density plot of the colour-magnitude Diagram with confident main sequence stars overplotted with a -1.83 in [Fe/H], 11.35 kpc in distance (value taken from \citealt{malhan2022global}) mock isochrone in purple and non-main sequence stars in grey (top middle), metallicity sensitive photometric distance distribution as a continuous line and photometric distance distribution as a dashed line (top right) and density distribution of main sequence stars in stream coordinates with radial velocity members from \citet{ji2020southern} and narrow component stream track (bottom). The counts(*) for the Gaussian smoothed density plot is simply the number of stars per $0.4\times0.1$ $\text{deg}^2$. Note a discontinuity clearly visible at $\phi_1\sim15^{\circ}$}
    \label{fig:jhelum-distance}
\end{figure*}

\subsection{Jhelum stellar stream}\label{3.3}

Another stream that we see from the second panel of Fig. \ref{fig:binned-velocity-sky} is the Jhelum stellar stream. The Jhelum stream was discovered by \citet{shipp2018stellar} using the first three years of multi-band optical imaging data from the Dark Energy Survey (DES). Since then, the stream has been extensively studied mostly due to its complex morphology and several sub-components (for most recent studies see, \citetalias{woudenberg2022characterization},
\citealt{li2022s}). 

To pick up candidates belonging to Jhelum stellar stream, we follow the same procedure as before and draw a rough polygon around the velocity peaks in the Gaussian smoothed binned velocity moments in the sky for on-stream candidates as shown in the top panels of Fig. \ref{fig:jhelum-selection}. We randomly pick up the same amount of stars from all directions outside the stream polygon and dub them off-stream. Their respective proper motion density in (l,b) coordinates is shown in the bottom panels of Fig. \ref{fig:jhelum-selection}. We can clearly see an overdensity of proper motion around the four-sided grey polygon drawn in Fig. \ref{fig:jhelum-selection}. We select stars from this proper motion peak and use them as Jhelum stellar stream faint candidates for the rest of this subsection. 

To validate our selection, we fit an empirical polynomial, in a similar fashion with GD-1, on the proper motions in (RA, Dec) with respect to the stream coordinate $\phi_1$ defined by \citet{bonaca2019multiple}. We select all the candidates that satisfy the following two conditions based on the second degree polynomial fit on proper motion in (RA, Dec) with respect to $\phi_1$:

\begin{gather}
    \mu_\text{RA} \pm (\overline{\delta(\mu_\text{RA})}+3\sigma_{\delta(\mu_\text{RA})}) < -4.52\phi_1^2-1.34\phi_1+6.62\\
    \mu_\text{Dec} \pm (\overline{\delta(\mu_\text{Dec})}+3\sigma_{(\delta(\mu_\text{Dec}))}) < 2.22\phi_1^2-4.11\phi_1-4.62
	\label{eq:jhelum-pm}
\end{gather}

where $\phi_1$ is in radians and $\mu_\text{RA}$ and $\mu_\text{Dec}$ are in mas/yr. We allow here a $3\sigma$ deviation instead of $2\sigma$ owing to the complex multiple components in the stream that are debated to have similar \citep{bonaca2019multiple} or slightly different \citep{shipp2019proper} proper motion distribution. In the final distribution of stream star candidates, we find three components of Jhelum - the narrow and broad components almost parallel to each other reported by \citet{bonaca2019multiple} and the tertiary spur component reported after the advent of Gaia EDR3 recently in \citetalias{woudenberg2022characterization}. We fit the following three second degree polynomials to these sub-components by dividing the stream sky space into three parts based on their overdensities: 

\begin{gather}
    \label{eq:jhelum-sky1}
    \phi_2^1 = F_\text{narrow}(\phi_1) = 0.112\phi_1^2-0.017\phi_1+0.009\\
    \label{eq:jhelum-sky2}
    \phi_2^2 = F_\text{broad}(\phi_1) = 0.137\phi_1^2-0.004\phi_1-0.018\\
    \phi_2^3 = F_\text{spur}(\phi_1) = -0.099\phi_1^2-0.059\phi_1+0.039
\end{gather}

where $(\phi_1,\phi_2)$ are in radians. Equation (\ref{eq:jhelum-sky1}) fitted for the narrow component of Jhelum is similar to what is fitted by \citetalias{woudenberg2022characterization}. The polynomial fitting and confident member selection are illustrated in Fig. \ref{fig:jhelum-validation}. In the last panel of this figure, we also illustrate the fact that the Jhelum broad component is not just $\sim0.9^{\circ}$ shifted from the narrow component (thick dashed line proposed by \citealt{bonaca2019multiple}) but is better described by equation (\ref{eq:jhelum-sky2}).
These Jhelum stream confident members have no cross-matches with existing publicly available spectroscopic surveys mostly because all our candidates are extremely faint (G > 18.5) and in the southern sky. Therefore, to improve the distances sensitive to the stream's metallicity, we adopt the distribution based on the mean and standard deviation proposed by \citet{li2022s}. We fit a mock isochrone using \citetalias{Kim_2021} photometric metallicity grid at [Fe/H]=-1.83 and find the mean distance to be 10.44 kpc which is a bit closer than what is observed in the literature. This can be attributed to the fact that we only probe the nearby main sequence of the Jhelum stream due to Gaia's limiting magnitude. This distance fitting is illustrated in Fig. \ref{fig:jhelum-distance}. In the bottom panel of this figure, we plot the stream density of the main sequence stars. Radial velocity members from \citet{ji2020southern} are overplotted and we see that these members are consistent with our choice of stream track. We also plot the metallicities derived for these candidates in the top panel as a histogram, but we can see that all of these stars are more metal poor than the observed mean metallicity. This is because the stars observed with high-resolution spectroscopy in \citet{ji2020southern} were preferentially selected to be metal-poor compared to all possible Jhelum members. It is important to note that we see a density break in the middle of the narrow component around $\phi_1\sim15$ with one side of the stream going downwards and the other side upwards for almost $2^{\circ}$. These two tracks causing a density break are 0.15$^{\circ}$ and -0.38$^{\circ}$ away from the narrow stream track and the break in density is $\num{\sim1}\sigma$ below the density of the narrow stream. This was seen and simulated as the kink feature in \citetalias{woudenberg2022characterization} caused due to interactions with Sagittarius but not resolved as clearly as we see here. As such, this showcases an advantage of probing fainter counterparts of such complex stellar streams. 

\subsection{Lower proper motion tail of Sagittarius}\label{3.4}

\begin{figure}
    \centering
	\includegraphics[width=\columnwidth]{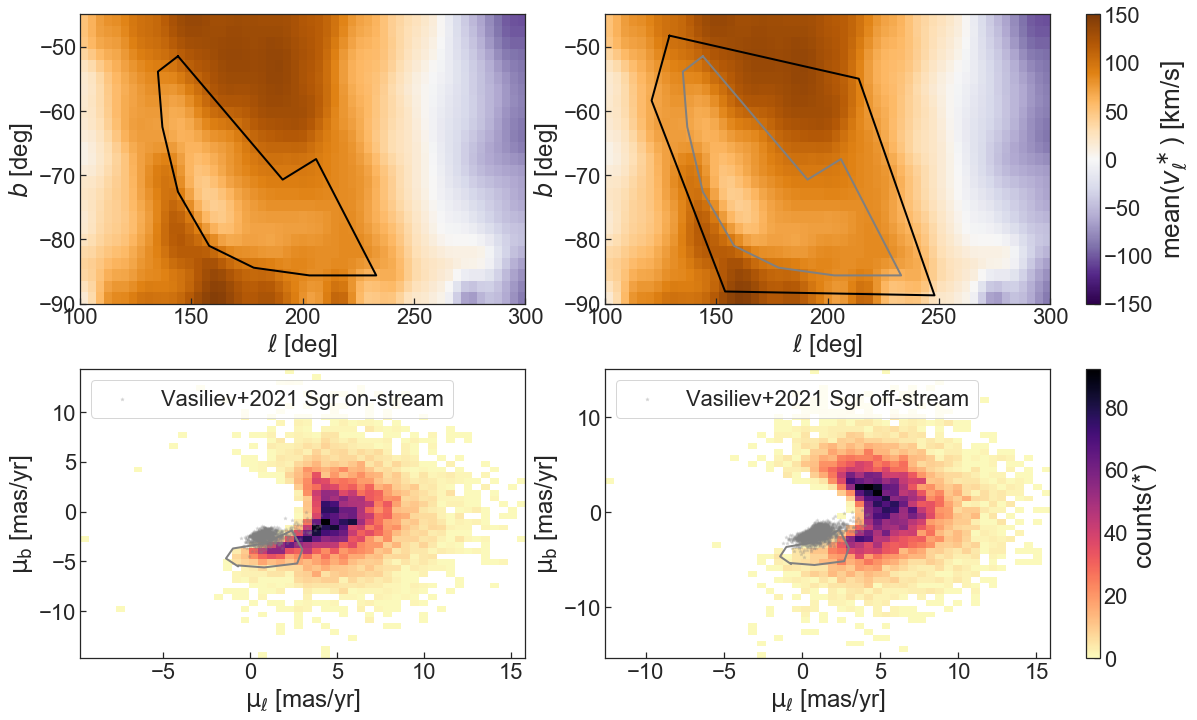}
    \caption{Binned longitudinal velocity moments in the sky at d>8 kpc with the (rough) polygon selection of a part of Sagittarius picked up by the RPM sample (top) and Gaia proper motions in (l,b) (bottom) within the polygon selection of the stream - on stream track (left) and outside the polygon selection of the stream - off stream track (right). This is overplotted with stars in the on-stream and off-stream polygon of the Sagittarius RGB sample from \citet{vasiliev2021tango}.}
    \label{fig:some-other-selection}
\end{figure}

\begin{figure}
    \centering
	\includegraphics[width=\columnwidth]{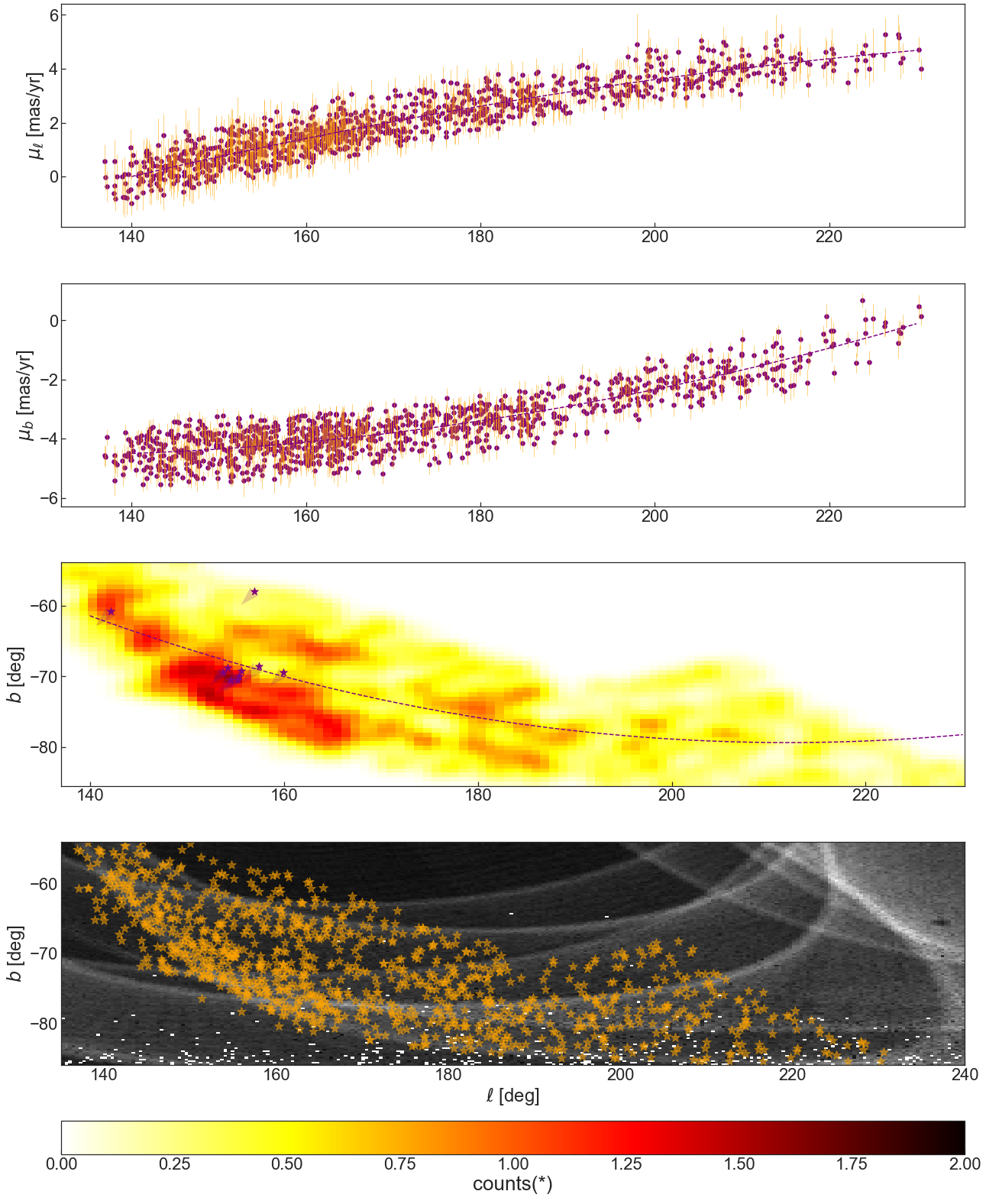}
    \caption{Proper motion in longitude direction (top), latitude direction (middle top) with respect to the Galactic longitude $\ell$ and density of candidate members on Galactic coordinates with radial velocity members as purple stars with longitudinal velocity quivers (middle bottom). Candidate members overplotted on the Gaia's scanning pattern in this region created using \texttt{astrometric\_n\_obs\_al} parameter (bottom). Dashed lines in plots one to three are second degree polynomial fits on the corresponding parameters. The counts(*) for the Gaussian smoothed density plot is simply the number of stars per $0.9\times0.64$ $\text{deg}^2$.}
    \label{fig:some-other-validation}
\end{figure}

\begin{figure*}
    \centering
	\includegraphics[width=\textwidth]{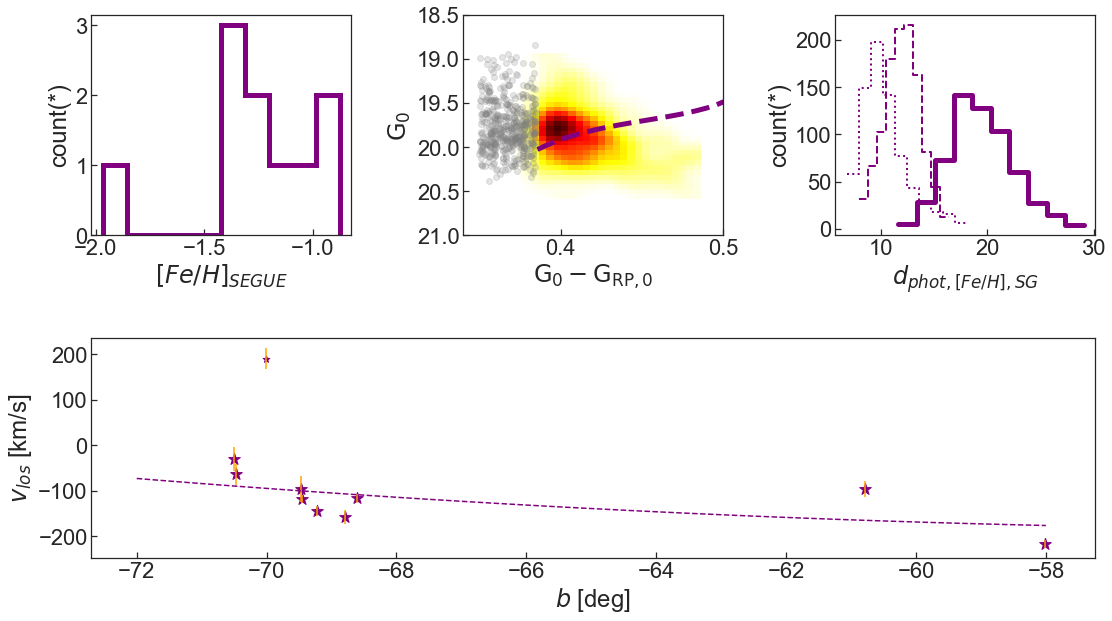}
    \caption{Metallicity distribution of radial velocity members (top left), colour-magnitude Diagram density distribution of confident main sequence with a -1.25 [Fe/H], 19.5 kpc (mean distance on the subgiant branch fit) MIST isochrone in purple and turnoff stars in grey (top middle), metallicity sensitive sub-giant photometric distance distribution as a continuous line, photometric distance distribution as a dashed line, and metallicity sensitive main-sequence turnoff photometric distance distribution as a dotted line (top right) and latitude versus line-of-sight velocity with a second degree polynomial fit on the track ignoring the one stream star contamination at $\sim$200 km/s (bottom).}
    \label{fig:some-other-distance}
\end{figure*}

\begin{figure}

    \centering
	\includegraphics[width=\columnwidth]{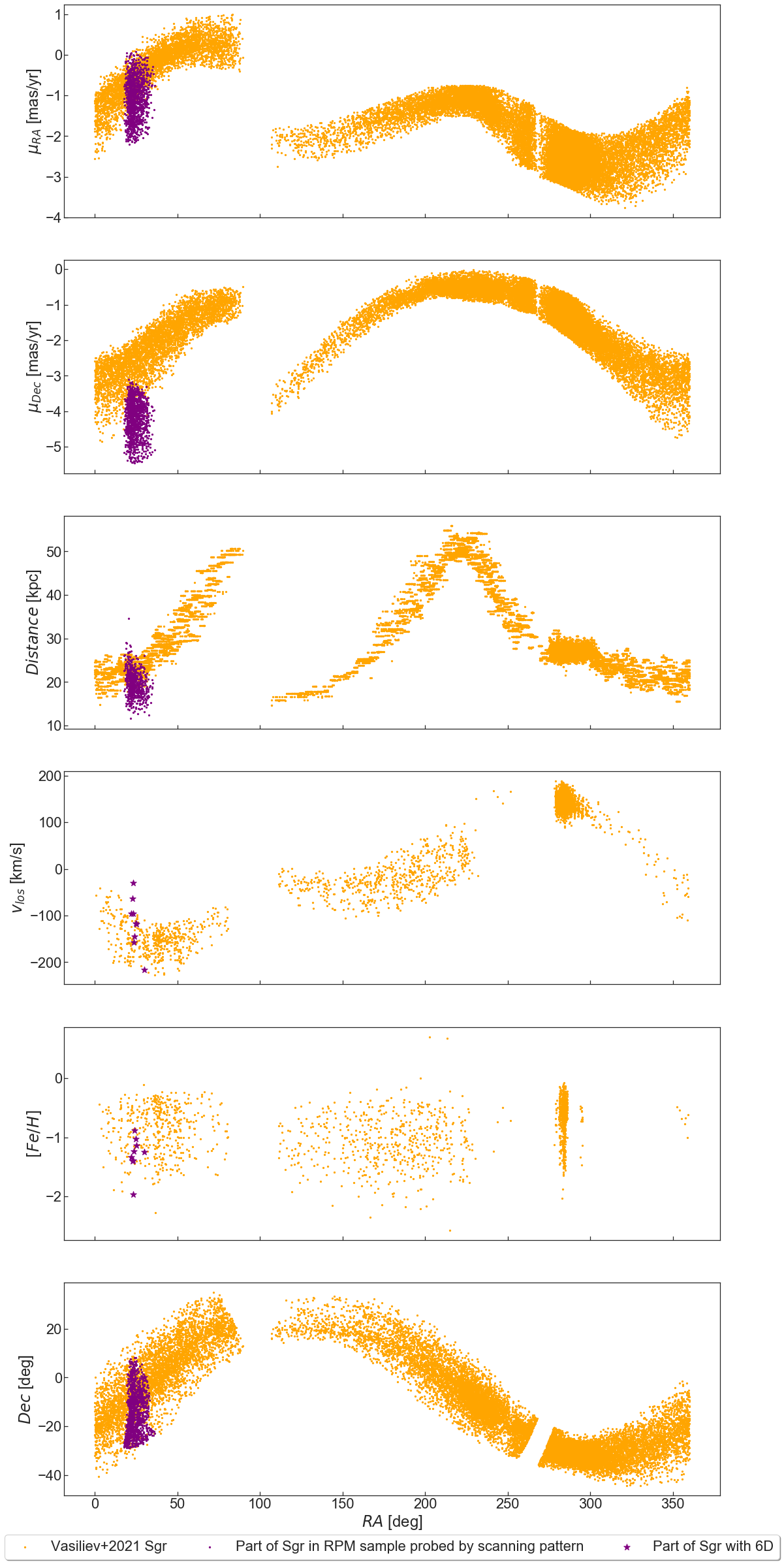}
    \caption{Observable properties of part of Sagittarius picked up by the RPM halo sample compared to \citet{vasiliev2021tango} RGB candidates belonging to the Sagittarius stream.}
    \label{fig:some-other-sag-coomparison}
\end{figure}

We pick up another clearly visible structure in Fig. \ref{fig:binned-velocity-sky}, top panel, around $l\sim150-200^\circ$ and $b<0^\circ$. Interestingly, we do not recognise this candidate stellar stream as a stream known in the literature. It overlaps with the Sagittarius broad stellar stream Additionally, it seems to be close to a great circle in the sky with the stellar streams Palca \citep{shipp2018stellar,li2022s} and Cetus \citep{chang2020ngc,yuan2022complexity} and overlaps significantly in the sky with the second structure almost parallel to the Cetus stream discovered by \citet{thomas2022cetus}, but it does not match these known streams and overdensities in proper motion space. To investigate this further, we perform a similar selection method drawing a rough polygon around the velocity peaks that are sufficiently different from the background halo and a control sample around the polygon on-stream selection. The density of proper motion in (l,b) shows a clear extension on the on-stream track at $(\mu_{\ell},\mu_{b})=(0,-5)$. This is illustrated in Fig. \ref{fig:some-other-selection}. It bears resemblance to the Sagittarius stream in certain observables, but extends to lower proper motions as well as lower distances (at $\sim$10 kpc) compared to Sagittarius samples defined in the literature so far. In the bottom panels of this figure, we overplot Sagitarrius RGB stream star candidates selected by \citet{vasiliev2021tango} within the on-stream and off-stream polygon. Sagittarius overlaps with the structure we see in this region and the proper motions, though do not entirely match. Rather, it looks like a lower end extension in the proper motion space of Sagittarius that is being picked up only at this region of the sky by the RPM sample. We notice that we lose the actual peak of Sagittarius proper motions because of the RPM sample's inherent selection bias against halo stars with small proper motions.

Similar to our procedure for GD-1 and Jhelum, we fit a second degree polynomial to the proper motion in (l,b) directions with respect to the longitude. The stream stars show a clear trend with increasing longitude. We also fit a second degree polynomial to the $(l,b)$ sky space. The fitted polynomial functions are as follows:
\begin{gather}
    \mu_{\ell} = F_1(\ell) = -0.87\ell^2+8.62\ell-15.85\\
    \mu_{b} = F_2(\ell) = 1.39\ell^2-6.17\ell+2.29\\
    b = F_3(\ell) = 0.19\ell^2-1.46\ell+1.32
\end{gather}
where $\ell$ and $b$ are in radians and $\mu_{\ell}$ and $\mu_b$ are in mas/yr. The polynomial fit on proper motion and on sky density distribution along with Gaia's scanning pattern around it is shown in Fig. \ref{fig:some-other-validation}. We see two sub-components in this structure selection that are of higher density than the rest. These two high density regions clearly overlap with two arcs in the background caused by Gaia's scanning pattern. Gaia scans this region more frequently than the surrounding regions which means that it probes deeper in these regions. This might explain why we seem to have picked up the lower proper motion part of Sagittarius only in this apparently stream-like feature. We do not see the same signal for the rest of the Sagittarius stream, because they are too far and deep for the sample to reach while Gaia's scanning pattern makes this part of Sagittarius pop out. Ten stars have metallicities and line-of-sight velocities after cross match with SDSS/SEGUE. Although these stars have poor signal-to-noise, it is good enough to validate that the metallicities fall into the range of metallicities of Sagittarius selected from \citet{vasiliev2021tango} in this region as it can be seen in the bottom panels of Fig. \ref{fig:some-other-sag-coomparison}. These stars are plotted in the sky as purple stars with velocity quivers in the third panel of Fig. \ref{fig:some-other-validation}. 

Finally, we attempt to improve the distances to the stars using the metallicity information from SEGUE cross-matches. We realise that a large majority of the stars in the feature we pick up are actually bluer than our preferred colour selection with $(G_0-G_\text{RP,0})<0.45$, which means that the distances are less reliable. It is possible for the turnoff stars higher on the turnoff (from the sub-giant side of the CMD) to bleed into our sample. We create a three dimensional interpolation for MIST isochrones (as they do work well in the sub-giant branch) using age, metallicity and Gaia colour as inputs. By feeding the mean metallicity of -1.25 (based on the 10 SEGUE cross-matches) and an age of 12.5 Gyr, we find that the two distance solutions possible either give a distance of $\sim$10 kpc, or $\sim$20 kpc, to the feature. The latter distance would correspond to the predicted Sagittarius stream stars in that region of the sky \citet{vasiliev2021tango} and would explain why the feature does not show any redder stars. The distance and metallicity distribution and CMD  for these stars are shown in the top panels of Fig. \ref{fig:some-other-distance}. In the bottom panel, we show the latitude versus line-of-sight velocities for the radial velocity members and fit a second degree polynomial ignoring the star at $\sim$200 km/s as contamination. This polynomial fit is described by the following equation:

\begin{gather}
    v_\text{los} = F(b) = 964.4b^2+1765.9b+623.2
\end{gather}

where $b$ is in radians and $v_\text{los}$ is in km/s. We compare the observable properties of these member stars with Sagittarius stream candidates selected by \citet{vasiliev2021tango}. These comparison figures are shown in Fig. \ref{fig:some-other-sag-coomparison}. The structure we see agrees with Sagittarius stream candidates in that region in almost all the observables, and extends it in proper motions. We conclude that this feature is indeed a part of Sagittarius and that it looks like a thin stream crossing Sagittarius due to a combination of Gaia's scanning law - providing more fainter stars in this feature - and the RPM sample selection effect on small proper motion. 

\subsection{Summary of stream properties}
The distance properties and the number of candidate members selected are summarised for all three discussed candidate streams in Table \ref{tab:streams}. The columns that will be provided as a part of the stream member candidate catalogue are shown as a part of Table \ref{tab:mem}.


\begin{table}
	\centering
	\caption{The properties of stream members selected using the binned velocity space from the reduced proper motion catalogue }
	\label{tab:streams}
	\begin{tabular}{cccccc} 
		\hline
		Name & $\text{N}$ & $\text{N}_\text{MS/SG}$ &$\text{d}_\text{phot}^\text{hc}$ & $\text{d}_\text{phot,[Fe/H]}^\text{hc}$ \\
		&&&[kpc]&[kpc]\\
		\hline
		GD-1 & 1155 & 783 & 8.82  & 7.84 & \\
		Jhelum & 1353 & 692 & 11.32  & 10.40 & \\
		Part of Sagittarius & 1114 & 589 & 12.03 & 19.50 & \\
		\hline
	\end{tabular}
\\Note: The full stream star candidate list will be made available online
\end{table}

\section{Conclusions and Outlook}\label{4}

With the advent of the European Space Agency's Gaia mission and its recent data releases EDR3 and DR3, the astronomy community has obtained the largest-ever cartograph of our Milky Way galaxy with unprecedented astrometric parameters. In this paper, we have presented a catalogue of $\sim$47 million halo stars on the main sequence with high tangential velocity selected only using Gaia DR3 proper motions and photometry. This is made possible using the reduced proper motion that when plotted versus Gaia colours mimics the colour-magnitude diagram for populations with different tangential velocities. Most of the literature methods to select halo stars make use of distance and/or spectroscopy information. One much used example of this, is to create a 'Toomre' diagram of velocities, where a cut can be made to separate stars that move fast with respect to the solar motion (e.g., $|V-V_\text{LSR}|>210$ km/s) to isolate halo stars \citep[see for implementations of such a cut for instance][]{bonaca2017gaia,koppelman2018one}. The disadvantage of such a method is however that it is limited to stars with good parallaxes and line-of-sight velocities. Instead, we build in this work a catalogue of inner halo stars out to $\sim$21 kpc. This sample is five times bigger with less systematics and much improved completeness beyond 5 kpc (thanks to Gaia DR3) than the original \citetalias{Koppelman_2021} RPM sample produced using Gaia DR2 and showcases the science potential of such a sample with the upcoming data releases of Gaia (for eg., Gaia DR4). The distance makes up an important aspect of the 6D information often used in the study of the dynamical evolution of the Milky Way halo. Here, we calculate photometric distances to these stars with simple linear colour-magnitude relation for these stars on the main sequence. The typical uncertainty on these derived photometric distances is $\sim$7$\%$ which is more reliable and probes farther away than would be possible using Gaia parallaxes.

We explore the possibility of using the binned velocity space of this dataset at relatively higher distances to independently pick up candidate members and main sequence stars belonging to three example streams. The use of main-sequence stars, rather than brighter giants, allows us to trace low surface brightness counterparts. Two cold streams with complex morphology: GD-1 and Jhelum are picked up by the sample and explored in detail. 

We fit polynomial functions for the observable properties of the stream members with respect to the main stream coordinate $\phi_1$. We see gaps and overdensities similar to what is observed in the literature and often enhanced to a greater contrast. For GD-1, we see a density gap at $\phi_1\sim40^{\circ}$ which could be a probable position for the now fully disrupted GD-1's progenitor \citep{webb2019searching}. However, this is not conclusive, because the distribution of the tracks causing the density breaks overlap quite a lot. More member stars and radial velocities for these members should be helpful for further analysis. We also see a similar density gap in Jhelum at $\phi_1\sim15^\circ$ resolved very well due to improved Gaia astrometry and photometry. The several components of the stream was recently explored using N-body simulations as due to a tentative close interaction with Sagittarius \citepalias{woudenberg2022characterization} while some of these density breaks and variations still remain to be explained. We improve the photometric distances calculated by the entire RPM sample by folding in the metallicity distribution information for these stream members using photometric metallicity grids computed by \citet{Kim_2021}. Our selection method is validated using line-of-sight and metallicity information from cross-matches with spectroscopic surveys (for GD-1) or candidate members from the literature (for Jhelum) and returns very high success rates as $\sim90\%$ of the overlapping stars turn out to be true members.

We also see part of Sagittarius in the binned velocity space in a negative latitude direction with similar properties, but slightly more negative proper motions compared to literature samples for the stream. This structure is characterized better by fitting distances on the sub-giant branch using MIST isochrones and metallicity information from SEGUE cross-matches. Based on the available information, the structure is likely to belong to the Sagittarius stream which is visible in this catalogue only in that part of the sky due to a combination of the kinematic selection bias of this sample, its incompleteness at higher distances due to the intrinsic faintness of main sequence stars, and Gaia's scanning pattern. Future Gaia data releases may be less affected by Gaia's scanning pattern and more reliable to hunt for coherent streams on such binned velocity space. However, for this catalogue, this detection illustrates that care needs to be taken in the interpretation of substructures. 

Aiding future follow-up studies, we provide candidate member catalogue with our photometric distances, further refined by using the metallicity distribution of these streams as input. A more systematic exploration of nearby stellar streams by picking up proper motion peaks would further enhance the potential of this catalogue. As such, we can push the substructure searches to Gaia’s magnitude limits. Using this sample in pseudo 6D space by setting the line-of-sight velocity to zero in regions of the sky where the pseudo space mimics the real Galactic velocities (ie., near Galactic centre, anticentre and poles) combined with added metallicity information can be used in the detailed chemodynamical analysis of the inner stellar halo.  

\subsection{Synergies with other datasets}
Such a main sequence stars sample with the largest halo cartograph in the era of Gaia can be used to provide spectroscopic targets to the next big spectroscopic surveys such as WEAVE \citep{dalton2012weave}, 4MOST \citep{de20194most}, and SDSS-V \citep{kollmeier2019sdss} because this catalogue complements the Gaia DR3 source spectra at the fainter end. Even low-resolution spectroscopic follow-up that can provide us with the missing line-of-sight velocities and/or metallicities can be extremely useful to disentangle the merger history of the inner stellar halo. In a subsequent paper, we intend to explore the cross-match of this sample with photometric metallicity information from the highly metallicity Pristine narrow-band survey \citep{starkenburg2017pristine} in the Northern Hemisphere ($\sim$1 million sources in overlap). While still lacking line-of-sight information, such a cross-match does provide us with valuable information about the metallicity structure of the halo at different distances and in various directions \citep[see e.g.,][]{youakim2020pristine}. Additionally, the metallicity information from Pristine allows us to improve on the derived distances, as we have showcased in this work by folding in spectroscopically known metallicities for the streams, thereby avoidingdistance biases. This effort will be made possible with this sub-sample. The sample presented in this work already allows a deeper (and fainter) dive into studying the structure of the Milky way inner halo, while holding even greater promise for unravelling the complex formation history of our Galaxy with future Gaia data releases. It is a golden age to do Galactic Archaeology.

\section*{Acknowledgements}

The authors thank the referee, Dr Ricardo Schiavon for their constructive comments and suggestions that helped improve the text and the figures. AV thanks Hanneke Woudenberg, Tomás Ruiz-Lara, Francisco Ardevol Martinez and Ewoud Wempe for their helpful discussions. AV thanks Dr Bokyoung Kim for providing access to their nearby reduced proper motion sample. AV and ES benefited from support from the International Space Science Institute (ISSI) in Bern, CH, thanks to the funding of the Team “the early Milky Way”. ES acknowledges funding through VIDI grant "Pushing Galactic Archaeology to its limits" (with project number VI.Vidi.193.093) which is funded by the Dutch Research Council (NWO). AH gratefully acknowledge financial support from a Spinoza prize from the Dutch Research Council
(NWO) and HHK gratefully acknowledges financial support from the Martin A.
and Helen Chooljian Membership at the Institute for Advanced Study.

This work has made use of data from the European Space Agency (ESA) mission {\it Gaia} (\url{https://www.cosmos.esa.int/gaia}), processed by the {\it Gaia} Data Processing and Analysis Consortium (DPAC, \url{https://www.cosmos.esa.int/web/gaia/dpac/consortium}). Funding for the DPAC has been provided by national institutions, in particular the institutions participating in the {\it Gaia} Multilateral Agreement. 

This work also uses SDSS SEGUE data. Funding for SDSS-III has been provided by the Alfred P. Sloan Foundation, the Participating Institutions, the National Science Foundation, and the U.S. Department of Energy Office of Science. The SDSS-III web site is http://www.sdss3.org/.
SDSS-III is managed by the Astrophysical Research Consortium for the Participating Institutions of the SDSS-III Collaboration including the University of Arizona, the Brazilian Participation Group, Brookhaven National Laboratory, Carnegie Mellon University, University of Florida, the French Participation Group, the German Participation Group, Harvard University, the Instituto de Astrofisica de Canarias, the Michigan State/Notre Dame/JINA Participation Group, Johns Hopkins University, Lawrence Berkeley National Laboratory, Max Planck Institute for Astrophysics, Max Planck Institute for Extraterrestrial Physics, New Mexico State University, New York University, Ohio State University, Pennsylvania State University, University of Portsmouth, Princeton University, the Spanish Participation Group, University of Tokyo, University of Utah, Vanderbilt University, University of Virginia, University of Washington, and Yale University.

Throughout this work, we’ve made use of the following packages and tools: \texttt{vaex} \citep{breddels2018vaex}, \texttt{NumPy} \citep{oliphant2006guide} \citep{van2011numpy}, \texttt{SciPy} \citep{scipy}, \texttt{matplotlib} \citep{hunter2007matplotlib}, \texttt{seaborn} \citep{seaborn}, \texttt{JupyterLab} \citep{kluyver2016jupyter}, \texttt{gala} \citep{price2017gala}, and \texttt{topcat} \citep{taylor2018topcat}

\section*{Data Availability}
All data used in this study are publicly available. The \texttt{PYTHON} codes used to create this catalogue is available here: \url{https://github.com/astroakshara/RPM-Catalogue-Gaia-DR3} and the full RPM halo catalogue along with the stream member candidates are available in this article and in its online supplementary material.



\bibliographystyle{mnras}
\bibliography{example} 

\begin{thebibliography}{}
\makeatletter
\relax
\def\mn@urlcharsother{\let\do\@makeother \do\$\do\&\do\#\do\^\do\_\do\%\do\~}
\def\mn@doi{\begingroup\mn@urlcharsother \@ifnextchar [ {\mn@doi@}
  {\mn@doi@[]}}
\def\mn@doi@[#1]#2{\def\@tempa{#1}\ifx\@tempa\@empty \href
  {http://dx.doi.org/#2} {doi:#2}\else \href {http://dx.doi.org/#2} {#1}\fi
  \endgroup}
\def\mn@eprint#1#2{\mn@eprint@#1:#2::\@nil}
\def\mn@eprint@arXiv#1{\href {http://arxiv.org/abs/#1} {{\tt arXiv:#1}}}
\def\mn@eprint@dblp#1{\href {http://dblp.uni-trier.de/rec/bibtex/#1.xml}
  {dblp:#1}}
\def\mn@eprint@#1:#2:#3:#4\@nil{\def\@tempa {#1}\def\@tempb {#2}\def\@tempc
  {#3}\ifx \@tempc \@empty \let \@tempc \@tempb \let \@tempb \@tempa \fi \ifx
  \@tempb \@empty \def\@tempb {arXiv}\fi \@ifundefined
  {mn@eprint@\@tempb}{\@tempb:\@tempc}{\expandafter \expandafter \csname
  mn@eprint@\@tempb\endcsname \expandafter{\@tempc}}}

\bibitem[\protect\citeauthoryear{Ahumada et~al.,}{Ahumada
  et~al.}{2020}]{ahumada202016th}
Ahumada R.,  et~al., 2020, The Astrophysical Journal Supplement Series, 249, 3

\bibitem[\protect\citeauthoryear{Andrae et~al.,}{Andrae
  et~al.}{2022}]{andrae2022gaia}
Andrae R.,  et~al., 2022, arXiv preprint arXiv:2206.06138

\bibitem[\protect\citeauthoryear{Angeli et~al.,}{Angeli
  et~al.}{2022}]{deangeli2022gaia}
Angeli F.~D.,  et~al., 2022, Gaia Data Release 3: Processing and validation of
  BP/RP low-resolution spectral data (\mn@eprint {arXiv} {2206.06143})

\bibitem[\protect\citeauthoryear{Arce \& Goodman}{Arce \&
  Goodman}{1999}]{arce1999measuring}
Arce H.~G.,  Goodman A.~A.,  1999, The Astrophysical Journal Letters, 512, L135

\bibitem[\protect\citeauthoryear{Balbinot, Cabrera-Ziri  \& Lardo}{Balbinot
  et~al.}{2021}]{balbinot2021}
Balbinot E.,  Cabrera-Ziri I.,   Lardo C.,  2021, Evidence for C, O, Al and Mg
  variations in the GD-1 stellar stream, \mn@doi{10.48550/ARXIV.2111.12626},
  \url {https://arxiv.org/abs/2111.12626}

\bibitem[\protect\citeauthoryear{Banik, Bovy, Bertone, Erkal  \& de Boer}{Banik
  et~al.}{2021}]{banik2021evidence}
Banik N.,  Bovy J.,  Bertone G.,  Erkal D.,   de Boer T.,  2021, Monthly
  Notices of the Royal Astronomical Society, 502, 2364

\bibitem[\protect\citeauthoryear{Belokurov et~al.,}{Belokurov
  et~al.}{2006}]{belokurov2006field}
Belokurov V.,  et~al., 2006, The Astrophysical Journal, 642, L137

\bibitem[\protect\citeauthoryear{Belokurov, Erkal, Evans, Koposov  \&
  Deason}{Belokurov et~al.}{2018}]{belokurov2018co}
Belokurov V.,  Erkal D.,  Evans N.,  Koposov S.,   Deason A.,  2018, Monthly
  Notices of the Royal Astronomical Society, 478, 611

\bibitem[\protect\citeauthoryear{Binney et~al.,}{Binney
  et~al.}{2014}]{binney2014new}
Binney J.,  et~al., 2014, Monthly Notices of the Royal Astronomical Society,
  437, 351

\bibitem[\protect\citeauthoryear{Bonaca, Conroy, Wetzel, Hopkins  \&
  Kere{\v{s}}}{Bonaca et~al.}{2017}]{bonaca2017gaia}
Bonaca A.,  Conroy C.,  Wetzel A.,  Hopkins P.~F.,   Kere{\v{s}} D.,  2017, The
  Astrophysical Journal, 845, 101

\bibitem[\protect\citeauthoryear{Bonaca, Conroy, Price-Whelan  \& Hogg}{Bonaca
  et~al.}{2019}]{bonaca2019multiple}
Bonaca A.,  Conroy C.,  Price-Whelan A.~M.,   Hogg D.~W.,  2019, The
  Astrophysical Journal Letters, 881, L37

\bibitem[\protect\citeauthoryear{Bonaca et~al.,}{Bonaca
  et~al.}{2020}]{bonaca2020high}
Bonaca A.,  et~al., 2020, The Astrophysical Journal Letters, 892, L37

\bibitem[\protect\citeauthoryear{Breddels \& Veljanoski}{Breddels \&
  Veljanoski}{2018}]{breddels2018vaex}
Breddels M.~A.,  Veljanoski J.,  2018, Astronomy \& Astrophysics, 618, A13

\bibitem[\protect\citeauthoryear{Buder et~al.,}{Buder
  et~al.}{2021}]{buder2021galah+}
Buder S.,  et~al., 2021, Monthly Notices of the Royal Astronomical Society,
  506, 150

\bibitem[\protect\citeauthoryear{{Cardelli}, {Clayton}  \& {Mathis}}{{Cardelli}
  et~al.}{1989}]{extinction2}
{Cardelli} J.~A.,  {Clayton} G.~C.,   {Mathis} J.~S.,  1989, \mn@doi [\apj]
  {10.1086/167900}, \href
  {https://ui.adsabs.harvard.edu/abs/1989ApJ...345..245C} {345, 245}

\bibitem[\protect\citeauthoryear{Chandra et~al.,}{Chandra
  et~al.}{2022}]{chandra2022distant}
Chandra V.,  et~al., 2022, arXiv preprint arXiv:2212.00806

\bibitem[\protect\citeauthoryear{Chang, Yuan, Xue, Simion, Kang, Li, Zhao  \&
  Zhao}{Chang et~al.}{2020}]{chang2020ngc}
Chang J.,  Yuan Z.,  Xue X.-X.,  Simion I.~T.,  Kang X.,  Li T.~S.,  Zhao
  J.-K.,   Zhao G.,  2020, The Astrophysical Journal, 905, 100

\bibitem[\protect\citeauthoryear{Choi, Dotter, Conroy, Cantiello, Paxton  \&
  Johnson}{Choi et~al.}{2016}]{choi2016mesa}
Choi J.,  Dotter A.,  Conroy C.,  Cantiello M.,  Paxton B.,   Johnson B.~D.,
  2016, The Astrophysical Journal, 823, 102

\bibitem[\protect\citeauthoryear{Cui et~al.,}{Cui et~al.}{2012}]{cui2012large}
Cui X.-Q.,  et~al., 2012, Research in Astronomy and Astrophysics, 12, 1197

\bibitem[\protect\citeauthoryear{Dalton et~al.,}{Dalton
  et~al.}{2012}]{dalton2012weave}
Dalton G.,  et~al., 2012, in Ground-based and Airborne Instrumentation for
  Astronomy IV. pp 220--231

\bibitem[\protect\citeauthoryear{Deason, Belokurov  \& Evans}{Deason
  et~al.}{2011}]{deason2011milky}
Deason A.,  Belokurov V.,   Evans N.,  2011, Monthly Notices of the Royal
  Astronomical Society, 416, 2903

\bibitem[\protect\citeauthoryear{Deason, Belokurov, Koposov  \&
  Lancaster}{Deason et~al.}{2018}]{deason2018apocenter}
Deason A.~J.,  Belokurov V.,  Koposov S.~E.,   Lancaster L.,  2018, The
  Astrophysical Journal Letters, 862, L1

\bibitem[\protect\citeauthoryear{Di~Matteo, Haywood, Lehnert, Katz, Khoperskov,
  Snaith, Gómez  \& Robichon}{Di~Matteo et~al.}{2019}]{blue-red-sequence}
Di~Matteo P.,  Haywood M.,  Lehnert M.~D.,  Katz D.,  Khoperskov S.,  Snaith
  O.~N.,  Gómez A.,   Robichon N.,  2019, \mn@doi [Astronomy & Astrophysics]
  {10.1051/0004-6361/201834929}, 632, A4

\bibitem[\protect\citeauthoryear{Dillamore, Belokurov, Evans  \&
  Price-Whelan}{Dillamore et~al.}{2022}]{dillamore2022impact}
Dillamore A.~M.,  Belokurov V.,  Evans N.~W.,   Price-Whelan A.~M.,  2022,
  arXiv preprint arXiv:2205.13547

\bibitem[\protect\citeauthoryear{Doke \& Hattori}{Doke \&
  Hattori}{2022}]{doke2022probability}
Doke Y.,  Hattori K.,  2022, arXiv preprint arXiv:2203.15481

\bibitem[\protect\citeauthoryear{Fabricius et~al.,}{Fabricius
  et~al.}{2021b}]{ruwe}
Fabricius C.,  et~al., 2021b, \mn@doi [Astronomy & Astrophysics]
  {10.1051/0004-6361/202039834}, 649, A5

\bibitem[\protect\citeauthoryear{Fabricius et~al.,}{Fabricius
  et~al.}{2021a}]{fabricius2021gaia}
Fabricius C.,  et~al., 2021a, Astronomy \& Astrophysics, 649, A5

\bibitem[\protect\citeauthoryear{{GRAVITY Collaboration} et~al.,}{{GRAVITY
  Collaboration} et~al.}{2018}]{GRAVITY2018}
{GRAVITY Collaboration} et~al., 2018, \mn@doi [Astronomy \& Astrophysics]
  {10.1051/0004-6361/201834294}, 618, L10

\bibitem[\protect\citeauthoryear{{Gaia Collaboration} et~al.,}{{Gaia
  Collaboration} et~al.}{2018}]{2018}
{Gaia Collaboration} et~al., 2018, \mn@doi [Astronomy & Astrophysics]
  {10.1051/0004-6361/201833051}, 616, A1

\bibitem[\protect\citeauthoryear{{Gaia Collaboration} et~al.,}{{Gaia
  Collaboration} et~al.}{2021}]{2021}
{Gaia Collaboration} et~al., 2021, \mn@doi [Astronomy & Astrophysics]
  {10.1051/0004-6361/202039657e}, 650, C3

\bibitem[\protect\citeauthoryear{{Gaia Collaboration} et~al.,}{{Gaia
  Collaboration} et~al.}{2022}]{babusiaux2022gaia}
{Gaia Collaboration} et~al., 2022, arXiv preprint arXiv:2206.05989

\bibitem[\protect\citeauthoryear{Gould}{Gould}{2007}]{gould2007investigation}
Gould A.,  2007 (\mn@eprint {arXiv} {0708.1326})

\bibitem[\protect\citeauthoryear{Grillmair \& Dionatos}{Grillmair \&
  Dionatos}{2006}]{Grillmair_2006}
Grillmair C.~J.,  Dionatos O.,  2006, \mn@doi [The Astrophysical Journal]
  {10.1086/505111}, 643, L17

\bibitem[\protect\citeauthoryear{Hejazi, L{\'e}pine, Homeier, Rich  \&
  Shara}{Hejazi et~al.}{2020}]{hejazi2020chemical}
Hejazi N.,  L{\'e}pine S.,  Homeier D.,  Rich R.~M.,   Shara M.~M.,  2020, The
  Astronomical Journal, 159, 30

\bibitem[\protect\citeauthoryear{Helmi}{Helmi}{2008}]{helmi2008stellar}
Helmi A.,  2008, The Astronomy and Astrophysics Review, 15, 145

\bibitem[\protect\citeauthoryear{{Helmi}}{{Helmi}}{2020}]{helmi2020}
{Helmi} A.,  2020, \mn@doi [\araa] {10.1146/annurev-astro-032620-021917}, \href
  {https://ui.adsabs.harvard.edu/abs/2020ARA&A..58..205H} {58, 205}

\bibitem[\protect\citeauthoryear{Helmi \& White}{Helmi \&
  White}{1999}]{helmi1999building}
Helmi A.,  White S.~D.,  1999, Monthly Notices of the Royal Astronomical
  Society, 307, 495

\bibitem[\protect\citeauthoryear{Helmi, White  et~al.}{Helmi
  et~al.}{1999}]{helmi1999debris}
Helmi A.,  White S.~D.,   et~al., 1999, Nature, 402, 53

\bibitem[\protect\citeauthoryear{Helmi et~al.}{Helmi
  et~al.}{2000}]{helmi2000mapping}
Helmi A.,  et~al., 2000, Monthly Notices of the Royal Astronomical Society,
  319, 657

\bibitem[\protect\citeauthoryear{Helmi, Babusiaux, Koppelman, Massari,
  Veljanoski  \& Brown}{Helmi et~al.}{2018}]{helmi2018merger}
Helmi A.,  Babusiaux C.,  Koppelman H.~H.,  Massari D.,  Veljanoski J.,   Brown
  A.~G.,  2018, Nature, 563, 85

\bibitem[\protect\citeauthoryear{Hunter}{Hunter}{2007}]{hunter2007matplotlib}
Hunter J.~D.,  2007, Computing in science \& engineering, 9, 90

\bibitem[\protect\citeauthoryear{Ibata, Gilmore  \& Irwin}{Ibata
  et~al.}{1995}]{ibata1995sagittarius}
Ibata R.~A.,  Gilmore G.,   Irwin M.~J.,  1995, Monthly Notices of the Royal
  Astronomical Society, 277, 781

\bibitem[\protect\citeauthoryear{Ibata, Thomas, Famaey, Malhan, Martin  \&
  Monari}{Ibata et~al.}{2020}]{ibata2020detection}
Ibata R.,  Thomas G.,  Famaey B.,  Malhan K.,  Martin N.,   Monari G.,  2020,
  The Astrophysical Journal, 891, 161

\bibitem[\protect\citeauthoryear{Ibata et~al.,}{Ibata
  et~al.}{2021}]{ibata2021charting}
Ibata R.,  et~al., 2021, The Astrophysical Journal, 914, 123

\bibitem[\protect\citeauthoryear{Ji et~al.,}{Ji et~al.}{2020}]{ji2020southern}
Ji A.~P.,  et~al., 2020, The Astronomical Journal, 160, 181

\bibitem[\protect\citeauthoryear{{Jones}}{{Jones}}{1972}]{1972ApJ...173..671J}
{Jones} E.~M.,  1972, \mn@doi [\apj] {10.1086/151454}, \href
  {https://ui.adsabs.harvard.edu/abs/1972ApJ...173..671J} {173, 671}

\bibitem[\protect\citeauthoryear{Jones, Oliphant, Peterson  et~al.}{Jones
  et~al.}{2001}]{scipy}
Jones E.,  Oliphant T.,  Peterson P.,   et~al., 2001, {SciPy}: Open source
  scientific tools for {Python}, \url {http://www.scipy.org/}

\bibitem[\protect\citeauthoryear{Juri{\'c} et~al.,}{Juri{\'c}
  et~al.}{2008}]{juric2008milky}
Juri{\'c} M.,  et~al., 2008, The Astrophysical Journal, 673, 864

\bibitem[\protect\citeauthoryear{Katz et~al.,}{Katz
  et~al.}{2022}]{katz2022gaia}
Katz D.,  et~al., 2022, Gaia Data Release 3 Properties and validation of the
  radial velocities (\mn@eprint {arXiv} {2206.05902})

\bibitem[\protect\citeauthoryear{Kim \& L{\'{e}pine}}{Kim \&
  L{\'{e}pine}}{2021}]{Kim_2021}
Kim L{\'{e}pine} 2021, \mn@doi [Monthly Notices of the Royal Astronomical
  Society] {10.1093/mnras/stab3671}, 510, 4308

\bibitem[\protect\citeauthoryear{{Kim} \& {L{\'e}pine}}{{Kim} \&
  {L{\'e}pine}}{2022}]{2022MNRAS.515..795K}
{Kim} B.,  {L{\'e}pine} S.,  2022, \mn@doi [\mnras] {10.1093/mnras/stac1794},
  \href {https://ui.adsabs.harvard.edu/abs/2022MNRAS.515..795K} {515, 795}

\bibitem[\protect\citeauthoryear{Kluyver et~al.,}{Kluyver
  et~al.}{2016}]{kluyver2016jupyter}
Kluyver T.,  et~al., 2016, Jupyter Notebooks-a publishing format for
  reproducible computational workflows..
IOS Press

\bibitem[\protect\citeauthoryear{Kollmeier et~al.,}{Kollmeier
  et~al.}{2019}]{kollmeier2019sdss}
Kollmeier J.,  et~al., 2019, Bulletin of the American Astronomical Society

\bibitem[\protect\citeauthoryear{Koposov, Rix  \& Hogg}{Koposov
  et~al.}{2010}]{koposov2010constraining}
Koposov S.~E.,  Rix H.-W.,   Hogg D.~W.,  2010, The Astrophysical Journal, 712,
  260

\bibitem[\protect\citeauthoryear{Koppelman \& Helmi}{Koppelman \&
  Helmi}{2021a}]{Koppelman_2021}
Koppelman H.~H.,  Helmi A.,  2021a, \mn@doi [Astronomy & Astrophysics]
  {10.1051/0004-6361/202038178}, 645, A69

\bibitem[\protect\citeauthoryear{Koppelman \& Helmi}{Koppelman \&
  Helmi}{2021b}]{koppelman2021determination}
Koppelman H.~H.,  Helmi A.,  2021b, Astronomy \& Astrophysics, 649, A136

\bibitem[\protect\citeauthoryear{Koppelman, Helmi  \& Veljanoski}{Koppelman
  et~al.}{2018}]{koppelman2018one}
Koppelman H.,  Helmi A.,   Veljanoski J.,  2018, The Astrophysical Journal
  Letters, 860, L11

\bibitem[\protect\citeauthoryear{Koppelman, Helmi, Massari, Price-Whelan  \&
  Starkenburg}{Koppelman et~al.}{2019}]{koppelman2019multiple}
Koppelman H.~H.,  Helmi A.,  Massari D.,  Price-Whelan A.~M.,   Starkenburg
  T.~K.,  2019, Astronomy \& Astrophysics, 631, L9

\bibitem[\protect\citeauthoryear{Li et~al.,}{Li et~al.}{2022}]{li2022s}
Li T.~S.,  et~al., 2022, The Astrophysical Journal, 928, 30

\bibitem[\protect\citeauthoryear{Lindegren et~al.,}{Lindegren
  et~al.}{2021}]{Parallax}
Lindegren L.,  et~al., 2021, \mn@doi [Astronomy & Astrophysics]
  {10.1051/0004-6361/202039653}, 649, A4

\bibitem[\protect\citeauthoryear{Majewski et~al.,}{Majewski
  et~al.}{2017}]{majewski2017apache}
Majewski S.~R.,  et~al., 2017, The Astronomical Journal, 154, 94

\bibitem[\protect\citeauthoryear{Malhan, Ibata  \& Martin}{Malhan
  et~al.}{2018}]{malhan2018ghostly}
Malhan K.,  Ibata R.~A.,   Martin N.~F.,  2018, Monthly Notices of the Royal
  Astronomical Society, 481, 3442

\bibitem[\protect\citeauthoryear{Malhan et~al.,}{Malhan
  et~al.}{2022}]{malhan2022global}
Malhan K.,  et~al., 2022, The Astrophysical Journal, 926, 107

\bibitem[\protect\citeauthoryear{{Marigo} et~al.,}{{Marigo}
  et~al.}{2017}]{parsec-isochrone}
{Marigo} P.,  et~al., 2017, \mn@doi [\apj] {10.3847/1538-4357/835/1/77}, \href
  {https://ui.adsabs.harvard.edu/abs/2017ApJ...835...77M} {835, 77}

\bibitem[\protect\citeauthoryear{Martin et~al.,}{Martin
  et~al.}{2022}]{martin2022pristine}
Martin N.~F.,  et~al., 2022, arXiv preprint arXiv:2201.01310

\bibitem[\protect\citeauthoryear{Mateu}{Mateu}{2022}]{mateu2022galstreams}
Mateu C.,  2022, arXiv preprint arXiv:2204.10326

\bibitem[\protect\citeauthoryear{McMillan}{McMillan}{2016}]{mcmillan2016mass}
McMillan P.~J.,  2016, Monthly Notices of the Royal Astronomical Society, p.
  stw2759

\bibitem[\protect\citeauthoryear{Myeong, Vasiliev, Iorio, Evans  \&
  Belokurov}{Myeong et~al.}{2019}]{myeong2019evidence}
Myeong G.,  Vasiliev E.,  Iorio G.,  Evans N.,   Belokurov V.,  2019, Monthly
  Notices of the Royal Astronomical Society, 488, 1235

\bibitem[\protect\citeauthoryear{Naidu, Conroy, Bonaca, Johnson, Ting,
  Caldwell, Zaritsky  \& Cargile}{Naidu et~al.}{2020}]{naidu2020evidence}
Naidu R.~P.,  Conroy C.,  Bonaca A.,  Johnson B.~D.,  Ting Y.-S.,  Caldwell N.,
   Zaritsky D.,   Cargile P.~A.,  2020, The Astrophysical Journal, 901, 48

\bibitem[\protect\citeauthoryear{Newberg, Yanny  \& Willett}{Newberg
  et~al.}{2009}]{newberg2009discovery}
Newberg H.~J.,  Yanny B.,   Willett B.~A.,  2009, The Astrophysical Journal,
  700, L61

\bibitem[\protect\citeauthoryear{{O'Donnell}}{{O'Donnell}}{1994}]{extinction1}
{O'Donnell} J.~E.,  1994, \mn@doi [\apj] {10.1086/173713}, \href
  {https://ui.adsabs.harvard.edu/abs/1994ApJ...422..158O} {422, 158}

\bibitem[\protect\citeauthoryear{Oliphant}{Oliphant}{2006}]{oliphant2006guide}
Oliphant T.~E.,  2006, A guide to NumPy.
Trelgol Publishing USA

\bibitem[\protect\citeauthoryear{Price-Whelan}{Price-Whelan}{2017}]{price2017gala}
Price-Whelan A.~M.,  2017, Journal of Open Source Software, 2, 388

\bibitem[\protect\citeauthoryear{Price-Whelan \& Bonaca}{Price-Whelan \&
  Bonaca}{2018}]{price2018off}
Price-Whelan A.~M.,  Bonaca A.,  2018, The Astrophysical Journal Letters, 863,
  L20

\bibitem[\protect\citeauthoryear{Recio-Blanco et~al.,}{Recio-Blanco
  et~al.}{2022}]{recio2022gaia}
Recio-Blanco A.,  et~al., 2022, arXiv preprint arXiv:2206.05534

\bibitem[\protect\citeauthoryear{Riello et~al.,}{Riello
  et~al.}{2021b}]{excess_flux}
Riello M.,  et~al., 2021b, \mn@doi [Astronomy & Astrophysics]
  {10.1051/0004-6361/202039587}, 649, A3

\bibitem[\protect\citeauthoryear{Riello et~al.,}{Riello
  et~al.}{2021a}]{riello2021gaia}
Riello M.,  et~al., 2021a, Astronomy \& Astrophysics, 649, A3

\bibitem[\protect\citeauthoryear{Ruiz-Lara, Matsuno, L{\"o}vdal, Helmi, Dodd
  \& Koppelman}{Ruiz-Lara et~al.}{2022}]{ruiz2022substructure}
Ruiz-Lara T.,  Matsuno T.,  L{\"o}vdal S.~S.,  Helmi A.,  Dodd E.,   Koppelman
  H.~H.,  2022, arXiv preprint arXiv:2201.02405

\bibitem[\protect\citeauthoryear{Schlegel, Finkbeiner  \& Davis}{Schlegel
  et~al.}{2009}]{article}
Schlegel J.,  Finkbeiner D.,   Davis M.,  2009, \mn@doi [The Astrophysical
  Journal] {10.1086/305772}, 500, 525

\bibitem[\protect\citeauthoryear{Sch{\"o}nrich, Binney  \&
  Dehnen}{Sch{\"o}nrich et~al.}{2010}]{schonrich2010local}
Sch{\"o}nrich R.,  Binney J.,   Dehnen W.,  2010, Monthly Notices of the Royal
  Astronomical Society, 403, 1829

\bibitem[\protect\citeauthoryear{{Sharma}, {Bland-Hawthorn}, {Johnston}  \&
  {Binney}}{{Sharma} et~al.}{2011}]{Sharma}
{Sharma} S.,  {Bland-Hawthorn} J.,  {Johnston} K.~V.,   {Binney} J.,  2011,
  \mn@doi [\apj] {10.1088/0004-637X/730/1/3}, \href
  {https://ui.adsabs.harvard.edu/abs/2011ApJ...730....3S} {730, 3}

\bibitem[\protect\citeauthoryear{Shih, Buckley, Necib  \& Tamanas}{Shih
  et~al.}{2022}]{shih2022via}
Shih D.,  Buckley M.~R.,  Necib L.,   Tamanas J.,  2022, Monthly Notices of the
  Royal Astronomical Society, 509, 5992

\bibitem[\protect\citeauthoryear{Shipp et~al.,}{Shipp
  et~al.}{2018}]{shipp2018stellar}
Shipp N.,  et~al., 2018, The Astrophysical Journal, 862, 114

\bibitem[\protect\citeauthoryear{Shipp et~al.,}{Shipp
  et~al.}{2019}]{shipp2019proper}
Shipp N.,  et~al., 2019, The Astrophysical Journal, 885, 3

\bibitem[\protect\citeauthoryear{Smith et~al.,}{Smith
  et~al.}{2009}]{10.1111/j.1365-2966.2009.15391.x}
Smith M.~C.,  et~al., 2009, \mn@doi [Monthly Notices of the Royal Astronomical
  Society] {10.1111/j.1365-2966.2009.15391.x}, 399, 1223

\bibitem[\protect\citeauthoryear{Starkenburg et~al.,}{Starkenburg
  et~al.}{2017}]{starkenburg2017pristine}
Starkenburg E.,  et~al., 2017, Monthly Notices of the Royal Astronomical
  Society, 471, 2587

\bibitem[\protect\citeauthoryear{Starkenburg et~al.,}{Starkenburg
  et~al.}{2019}]{starkenburg2019pristine}
Starkenburg E.,  et~al., 2019, Monthly Notices of the Royal Astronomical
  Society, 490, 5757

\bibitem[\protect\citeauthoryear{Taylor}{Taylor}{2018}]{taylor2018topcat}
Taylor M.,  2018, arXiv preprint arXiv:1811.09480

\bibitem[\protect\citeauthoryear{Thomas \& Battaglia}{Thomas \&
  Battaglia}{2022}]{thomas2022cetus}
Thomas G.~F.,  Battaglia G.,  2022, Astronomy \& Astrophysics, 660, A29

\bibitem[\protect\citeauthoryear{Van Der~Walt, Colbert  \& Varoquaux}{Van
  Der~Walt et~al.}{2011}]{van2011numpy}
Van Der~Walt S.,  Colbert S.~C.,   Varoquaux G.,  2011, Computing in science \&
  engineering, 13, 22

\bibitem[\protect\citeauthoryear{Vasiliev, Belokurov  \& Erkal}{Vasiliev
  et~al.}{2021}]{vasiliev2021tango}
Vasiliev E.,  Belokurov V.,   Erkal D.,  2021, Monthly Notices of the Royal
  Astronomical Society, 501, 2279

\bibitem[\protect\citeauthoryear{Wang, Zhang, Xue, Huang, Liu, Zhang  \&
  Yang}{Wang et~al.}{2022}]{wang2022probing}
Wang F.,  Zhang H.,  Xue X.,  Huang Y.,  Liu G.,  Zhang L.,   Yang C.,  2022,
  Monthly Notices of the Royal Astronomical Society, 513, 1958

\bibitem[\protect\citeauthoryear{{Waskom} et~al.,}{{Waskom}
  et~al.}{2016}]{seaborn}
{Waskom} M.,  et~al., 2016, {Seaborn: V0.7.0 (January 2016)},
  \mn@doi{10.5281/zenodo.45133}

\bibitem[\protect\citeauthoryear{Webb \& Bovy}{Webb \&
  Bovy}{2019}]{webb2019searching}
Webb J.~J.,  Bovy J.,  2019, Monthly Notices of the Royal Astronomical Society,
  485, 5929

\bibitem[\protect\citeauthoryear{Woudenberg, Koop, Balbinot  \&
  Helmi}{Woudenberg et~al.}{2022}]{woudenberg2022characterization}
Woudenberg H.~C.,  Koop O.,  Balbinot E.,   Helmi A.,  2022, arXiv preprint
  arXiv:2202.02132

\bibitem[\protect\citeauthoryear{Yanny et~al.,}{Yanny
  et~al.}{2009}]{yanny2009segue}
Yanny B.,  et~al., 2009, The Astronomical Journal, 137, 4377

\bibitem[\protect\citeauthoryear{Youakim et~al.,}{Youakim
  et~al.}{2020}]{youakim2020pristine}
Youakim K.,  et~al., 2020, Monthly Notices of the Royal Astronomical Society,
  492, 4986

\bibitem[\protect\citeauthoryear{Yuan, Chang, Beers  \& Huang}{Yuan
  et~al.}{2020}]{yuan2020low}
Yuan Z.,  Chang J.,  Beers T.~C.,   Huang Y.,  2020, The Astrophysical Journal
  Letters, 898, L37

\bibitem[\protect\citeauthoryear{Yuan et~al.,}{Yuan
  et~al.}{2022}]{yuan2022complexity}
Yuan Z.,  et~al., 2022, The Astrophysical Journal, 930, 103

\bibitem[\protect\citeauthoryear{de Boer, Belokurov, Koposov, Ferrarese, Erkal,
  C{\^o}t{\'e}  \& Navarro}{de~Boer et~al.}{2018}]{de2018deeper}
de Boer T.,  Belokurov V.,  Koposov S.,  Ferrarese L.,  Erkal D.,  C{\^o}t{\'e}
  P.,   Navarro J.,  2018, Monthly Notices of the Royal Astronomical Society,
  477, 1893

\bibitem[\protect\citeauthoryear{de Boer, Erkal  \& Gieles}{de~Boer
  et~al.}{2020}]{de2020closer}
de Boer T.,  Erkal D.,   Gieles M.,  2020, Monthly Notices of the Royal
  Astronomical Society, 494, 5315

\bibitem[\protect\citeauthoryear{de Jong et~al.,}{de~Jong
  et~al.}{2019}]{de20194most}
de Jong R.~S.,  et~al., 2019, arXiv preprint arXiv:1903.02464

\makeatother
\end{thebibliography}





\bsp	
\label{lastpage}
\end{document}